\documentclass{aa}

\usepackage{ulem}
\usepackage{graphicx}
\usepackage{txfonts}
\usepackage{color}
\usepackage{empheq}
\usepackage{newtxtext}
\usepackage[T1]{fontenc}
\usepackage{lipsum}
\usepackage{subcaption}         
\usepackage{lscape}          
\usepackage{placeins} 
\usepackage{afterpage}

\usepackage{stackengine}
\stackMath
\usepackage{xcolor}
\usepackage{dcolumn}
\usepackage{bm}
\usepackage{physics}
\usepackage{comment}
\usepackage[hidelinks]{hyperref}
\usepackage{float}
\usepackage{soul}
\makeatletter
\renewcommand*\aa@pageof{, page \thepage{} of \pageref*{LastPage}}
\makeatother
\begin{document}

\title{Constraints on asteroid-mass primordial black holes in dwarf galaxies using Hubble Space Telescope photometry}

\author{Nicolas Esser\inst{1}\fnmsep\thanks{nicolas.esser@ulb.be}\and Carrie Filion\inst{2}\and Sven De Rijcke\inst{3}\and Nitya Kallivayalil\inst{4}\and Hannah Richstein\inst{4}\and Peter Tinyakov\inst{1}\and Rosemary F.G.~Wyse\inst{5}}

\institute{Service de Physique Th\'{e}orique, Universit\'{e} Libre de Bruxelles, Boulevard du Triomphe,  CP225, 1050 Brussels, Belgium
\and Center for Computational Astrophysics, Flatiron Institute,
162 Fifth Ave, New York, NY 10010, USA
\and Department of Physics and Astronomy, Ghent University, Krijgslaan 281, 9000 Ghent, Belgium 
\and Department of Astronomy, The University of Virginia, 530 McCormick Road, Charlottesville, VA 22904, USA
\and Department of Physics \& Astronomy, The Johns Hopkins University, Baltimore, MD 21218, USA}

\date{\today}

\authorrunning{Esser et al.}
\titlerunning{Constraints on PBHs in UFDs using HST}

\abstract{Primordial black holes (PBHs) in the asteroid-mass range remain a viable and, until now, unconstrained dark matter (DM) candidate. If such PBHs exist, they could be captured by stars in DM-dominated environments with low velocity dispersion, such as ultra-faint dwarf galaxies (UFDs). The capture probability increases with the stellar mass, and captured PBHs would rapidly destroy their host stars. As a result, the presence of PBHs in UFDs would alter their stellar mass functions. Using photometric observations of three UFDs from the Hubble Space Telescope, we show that it is unlikely that their mass functions have been significantly modified by PBHs, and we place constraints on the PBH abundance. In the UFD Triangulum II, PBHs around $10^{19}$\,g are excluded at the $2\sigma$ ($3\sigma$) level from constituting more than $\sim55\%$ ($\sim78\%$) of the DM, while the possibility that PBHs represent the entirety of the DM is excluded at the $3.7\sigma$ level.}\keywords{Methods: statistical -- techniques: photometric -- stars: luminosity function, mass function -- galaxies: dwarf -- galaxies: stellar content -- dark matter}

\maketitle

\section{Introduction}
\label{sec:introduction}

Primordial black holes (PBHs) are theoretical objects that may have formed during the early stages of the Universe. First proposed more than 50 years ago by \cite{Zeldovich} and \cite{Hawking}, these black holes can have any mass, and may offer an explanation for dark matter (DM). However, many constraints have been placed on their abundance -- typically expressed as the PBH fraction relative to the total DM $f_\text{PBH}=\Omega_{\text{PBH}}/\Omega_{\text{DM}}$ -- across different mass ranges \citep{Carr}. Among these, the 'asteroid-mass' window, spanning approximately from $10^{17}$ to $10^{23}$g, has received particular attention. Indeed, while black holes within this range are microscopic in size, they are sufficiently massive to evade Hawking radiation, making them particularly challenging to probe. Nonetheless, significant efforts have been made in recent years to constrain this window -- for example, by extending the evaporation limits to higher masses (see e.g. \cite{Laha_2020,Coogan_2021}), exploiting various lensing effects expected from such PBHs (see e.g. \cite{Fedderke_2025} and references therein), or studying their gravitational interactions with Solar System bodies as they pass through (see e.g. \cite{Tran_2024}).

In this work, we assess the viability of asteroid-mass PBHs as a DM candidate through another mechanism. We explore the scenario in which stars capture PBHs during their formation \citep{Esser1,Oncins}. A star 'infected' by a PBH can undergo rapid accretion, leading to the star's destruction and the formation of a subsolar-mass black hole. The impact of this destruction process on stellar populations is expected to be significant in dark-matter-dominated environments with low velocity dispersion, where the probability of destruction approaches unity, as in some ultra-faint dwarf galaxies (UFDs).

UFDs are low luminosity, low surface-brightness systems that host old, metal-poor stellar populations. Along with the Galactic centre, they are the systems with the highest inferred dark-matter density. Due to their faintness, even the most massive UFDs orbiting the Milky Way were only recently detected and identified (see e.g. \citealt{Belokurov2007} and the review by \citealt{Simon_2019}).

\cite{Esser2} showed that the likelihood of PBH capture by stars depends on the stellar mass, with more massive stars having a higher probability of capturing PBHs and, consequently, of being destroyed. Thus, the presence of PBHs should modify the stellar mass functions of their host galaxies. If the DM in UFDs were composed of asteroid-mass PBHs, a depletion of more massive stars in these galaxies would be expected, leading to a bottom-heavy mass function compared to the case of non-PBH DM. In \cite{Esser2}, the sensitivity of the mass function to this effect was investigated using synthetic data. It was shown that, in the absence of PBH-induced depletion, the statistics of the existing data could already provide constraints on the PBH fraction in DM, $f_\text{PBH}$, to below 100\%.

In this paper, we analyse ultra-deep photometric observations of stars in three local UFDs --- Reticulum II, Segue 1, and Triangulum II --- conducted with the Hubble Space Telescope's (HST) advanced camera for surveys (ACS). By fitting the data with standard models for the stellar mass function, modified to account for the star destruction process by PBHs, we constrain the abundance of PBHs in these galaxies. More specifically, we employ a 'control sample' strategy exploiting the fact that, although the three galaxies in our sample share similar characteristics, the impact of PBHs on the stellar mass function of Reticulum II is expected to be negligible even when $f_\text{PBH}=1$, in contrast to the other two galaxies (see Sec.~\ref{sec:meritfac}). Consequently, we use the parameters inferred from Reticulum II to construct priors for the remaining UFDs, which we then use to derive constraints on the PBH abundance. With this approach we exclude, at $3.7\sigma$ confidence level, PBHs with masses around $\sim 10^{19}$g as constituting the entirety of the DM.

The rest of this paper is organised as follows. In Sec. \ref{sec:IMF} we review the impact of PBHs on the stellar mass function. In Sec. \ref{sec:UFDs data} the relevant information and photometric data for our sample of galaxies is gathered. Sec. \ref{sec:Bayesian} describes the Bayesian procedure used to analyse the data given our mass function model. In Sec. \ref{sec:results} we show the results of the analysis and the constraints which they imply for $f_\text{PBH}$. Sec. \ref{sec:conclusion} contains concluding remarks.

\section{Impact of PBHs on the stellar mass function}
\label{sec:IMF}
\subsection{Initial mass function}
\label{sec:IMF_only}

Ultra-faint dwarf galaxies are old and relatively simple systems whose stars formed in only a short period of star formation around $\sim 13$~Gyr ago \citep{Sacchi_2021}, which resulted in a population of metal-poor stars with a very narrow spread of ages. Of this population, only stars with masses $M\lesssim 0.8M_\odot$ -- i.e. those with lifetimes exceeding the age of the Universe -- presently remain on the main sequence (barring complications from binary evolution), while heavier stars have either entered the giant branch or have already turned into compact objects. UFDs are also not expected to exhibit mass segregation, implying that dynamical effects, such as possible tidal interactions with the Milky Way over their history, have not altered their mass function \citep{Gennaro_2018a}. Therefore, for low-mass stars, the present-day stellar mass function is, in the absence of the PBH destruction effect, essentially the same as the mass function of stars at birth, i.e. the initial mass function (IMF).

As a key topic in stellar astrophysics, the IMF has been extensively studied over the past few decades (see e.g. the review of \citealt{Bastian2010}). Two common fitting functions used to describe the IMF are the broken power law (BPL), and the log-normal (LN) distribution. The BPL
is parametrised as follows,  
\begin{equation}
\dfrac{dN_i}{dM}(M,\alpha_1,\alpha_2)\propto
\begin{cases}
&M^{\alpha_1} \text{ for }0.08M_\odot\le M<0.5M_\odot\\
&kM^{\alpha_2} \text{ for }M\ge0.5M_\odot,
\end{cases}
\label{eq:powerlaw}
\end{equation}
with the continuity constant $k=(0.5M_\odot)^{(\alpha_1-\alpha_2)}$ and $\alpha_{1,2}$ the two power-law exponents. For $\alpha_1=-1.3$ and $\alpha_2=-2.3$, the standard \cite{Kroupa} IMF is recovered. The LN distribution is given by 
\begin{equation}
\dfrac{dN_i}{dM}(M,M_c,\sigma_{\text{LN}})\propto \dfrac{1}{M}\exp\left(-\dfrac{\left[\log_{10}(M/M_c)\right]^2}{2\sigma_{\text{LN}}^2}\right),
    \label{eq:LN}
\end{equation}
where $M_c$ is the characteristic mass  and $\sigma_{\text{LN}}$ is the width of the distribution. The single-star \cite{Chabrier} IMF is recovered for $M_c=0.08M_\odot$ and $\sigma_{\text{LN}}=0.69$.

\subsection{Star destruction by PBHs}

When a protostellar cloud contracts to form a star, it drags along the surrounding DM, creating a DM overdensity around the newly formed star. If the DM consists of asteroid-mass PBHs, this process results in an enhanced density of PBHs orbiting the star. Some of these PBHs may follow orbits that pass through the star. During each passage, these PBHs lose energy due to dynamical friction and accretion of stellar material, causing their orbits to gradually shrink. After 
sufficient encounters, their orbits become entirely confined within the star. At this stage, the accretion of stellar material becomes significantly more efficient, so that, assuming \cite{Bondi_1952} accretion regime, the star gets destroyed in a time (much) shorter than $\sim 10$~Gyr.

This process, first applied to the case of neutron stars and white dwarfs \citep{Capela_2012,Capela_2014}, was studied by \cite{Esser1} and \cite{Oncins} in the context of main-sequence stars. It was subsequently shown that the probability of a star capturing a PBH increases with the star's mass. Consequently, more massive stars are more prone to destruction by PBHs, influencing the stellar mass function.

We follow the formalism of \cite{Esser2} to model the effect of PBHs on the stellar population in the mass range $[10^{19},10^{21}]$g. The capture of PBH by a star is a random process that can be characterised by the mean number $\bar{\mathcal{N}}$ of PBHs captured over a star's lifetime ($13$~Gyr for stars in UFDs). This quantity is proportional to the PBH fraction in DM, $f_\text{PBH}$, and to the local DM density, $\rho_\text{DM}$, and inversely proportional to the cube of the PBH velocity dispersion\footnote{Throughout this paper we use $\sigma_v$ to denote the 3D velocity dispersion, which is related to the measured line-of-sight velocity dispersion given by $\sigma_\text{los}=\sigma_v/\sqrt{3}$.}, $\sigma_v$. We assume that DM and stars follow the same isotropic velocity distribution. Given that both stars and DM are collisionless and reside within the same gravitational potential, this assumption is reasonable. Moreover, what is relevant for our analysis is the flat, low-velocity region of the velocity distribution, which is largely insensitive to the distribution’s tails that may differ between the two populations.

The mean captured number can be written as
\begin{equation}
\bar{\mathcal{N}} = f_\text{PBH}\,\eta\, \nu(M),
\end{equation}
where the 'merit factor', 
\begin{equation}
\eta = \frac{\rho_\text{DM}}{100~\text{GeV/cm}^3} \left( 
\frac{7~\text{km/s}}{\sqrt{2}\sigma_v} \right)^3
\label{eq:merit_factor},
\end{equation}
reflects the specific conditions in a given UFD. The function $\nu(M)$ represents the mean
number of PBHs captured under reference conditions: $f_\text{PBH}=1$, $\rho_\text{DM}=100~{\text{GeV}/\text{cm}^ 3}$ and $\sigma_v=7~{\text{km}/\text{s}}$, by a star at rest. The factor $\sqrt{2}$ accounts for the motion of stars relative to the DM halo.
The function $\nu(M)$ has been calculated numerically from theoretical arguments in \cite{Esser2}. 
For practical use, an empirical fit, valid in the stellar mass range $[0.2, 0.8] M_\odot$ is given by
\begin{equation}
    \nu(M)=a(M/M_\odot)^b+c,\label{eq:nu}
\end{equation}
with $a=3.8$, $b=0.69$ and $c=-0.88$.

According to Poisson statistics, the likelihood that stars avoid destruction by PBHs is given by 
\begin{equation}
\label{eq:surviprob}
    P_S(M,f_\text{PBH})=\exp(-\bar{\mathcal{N}}) = \exp[-f_\text{PBH}\eta\nu(M)].
\end{equation}
Note that this survival probability depends exponentially on $\eta$. Moreover, the dependence on $\eta$ is degenerate with that on $f_\text{PBH}$. Examples of the function $P_S$ are displayed in Fig.~\ref{fig:surviprob}.

It is usually assumed that stellar populations in UFDs are of a single age. Given the linear dependence of $\nu(M)$ on the time since star formation \citep{Tinyakov2024}, this assumption is expected to be a good approximation for our purposes. Under this single-age approximation, the present-day stellar mass function combines the IMF with the effect of PBHs, and is given by
\begin{equation}
\label{eq:PDMF}
    \frac{dN}{dM}(M,\theta,f_\text{PBH})\propto\frac{dN_i(M,\theta)}{dM}\times P_S(M,f_\text{PBH}),
\end{equation}
where $\theta$ stands for the IMF parameters, i.e. $\{\alpha_1,\alpha_2\}$ for the BPL and $\{M_C,\sigma_{\text{LN}}\}$ for the LN distribution. This mass function serves as a model that we fit to real data in Sec.~\ref{sec:Bayesian}.

\subsection{PBHs and binary stars}

The formalism presented in \cite{Esser2} applies to isolated stars. However, a significant fraction of stars are in binary systems, so it is essential to evaluate the impact of companion stars on the PBH capture probability. Three-body interactions, which involve the primary star, its companion, and a PBH, could alter the trajectories of PBHs around the primary star. These interactions might cause some PBHs to be ejected from star-crossing orbits, while others could be drawn into such orbits. The net effect of companion stars is, therefore, not obvious and has not yet been quantified.

In the absence of a complete solution to the problem, the simplest approach is to disregard PBH capture in binary systems altogether. However, this assumption is overly conservative. 
The significant uncertainties in the binary fraction within UFDs ultimately prevent meaningful constraints from being placed on the PBH abundance. To make further progress, a more detailed investigation of the PBH capture process is required. As we show in Appendix \ref{app:binary}, for smaller PBH masses, the capture probability is dominated by PBHs with initially tighter orbits, which are less affected by perturbations from a companion star. Consequently, the effect of the companion star can be ignored, and Eq.~\eqref{eq:nu} remains valid as is, for PBH masses around $10^{19}$g.

\section{Ultra-faint dwarf galaxy sample}
\label{sec:UFDs data}

The IMF has been studied in detail by \cite{Filion_2022} and \cite{Filion_2024} in five UFDs: Reticulum II, Segue 1, Triangulum II, Ursa Major II, and Bo\"otes I. Among these, the first three are similar in age, metallicity, and velocity dispersion, making them a good sample for our study. Relevant information about these galaxies, along with the photometric data used in this analysis, is provided in the following subsections.

\subsection{The merit factors}
\label{sec:meritfac}

The merit factors $\eta$ of the UFDs are inferred from the measured quantities compiled by \cite{Simon_2019}. We first estimated the total mass of a given UFD, based on its measured projected half-light radius $R_{1/2}$ and line-of-sight velocity dispersion $\sigma_\text{los}$, which we plugged into the half-light mass estimator from \cite{Wolf_2010}. The DM density was obtained by dividing the half-light mass by the half-light volume, which was taken as the volume of the sphere of the deprojected half-light radius given by $r_{1/2}=4R_{1/2}/3$ \citep{Simon_2019}. We assumed that DM dominates the mass of the UFD. The merit factor can then be computed according to Eq.~\eqref{eq:merit_factor}. The relevant information and the resulting merit factors are compiled in Table~\ref{tab:data1}.
\begin{table}[ht]
\caption{\label{tab:data1}
Dynamical properties of dwarf galaxies used throughout this work.}
\renewcommand{\arraystretch}{1.5}
\begin{tabular}{|c|c|c|c|c|}
\hline
  & Ret. II & Seg. 1 & Tri. II & Refs. \\ \hline
 $R_{1/2}$ [pc] & $51$ & $24$ & $16$ & [1] \\ \hline
 $\sigma_v$ [km/s] & $5.7$ & $6.4$ & $<5.9$ & [2,3,4] \\ \hline
 $\rho_\text{DM}$ [$\text{GeV/cm}^3$] & $15$ & $85$ & $161$ & / \\ \hline
 $\bm{\eta}$ & $\bm{0.097}$ & $\bm{0.39}$ & $\bm{0.95}$ & / \\ \hline
\end{tabular}
\\
\tablefoot{Where only an upper limit on $\sigma_\text{los}$ is available, we conservatively use this limit as a value of $\sigma_\text{los}$. References: [1] \cite{Munoz}, [2] \cite{Simon_2015}, [3] \cite{Kirby}, and [4] \cite{Simon_2011}.}
\end{table}

We also considered the case where the DM is not uniformly distributed but follows a cuspy Navarro-Frenk-White profile \citep{NFW} and the stars a cored profile \citep{Plummer_1911} within the UFD. We find that it typically results in a mild $\mathcal{O}(10\%)$ enhancement in the average star destruction probability, which is dependent on the stellar mass. A more detailed investigation of non-uniform DM and stellar distributions is left for future work.

\begin{figure}[h!]
\includegraphics[width=1\columnwidth]{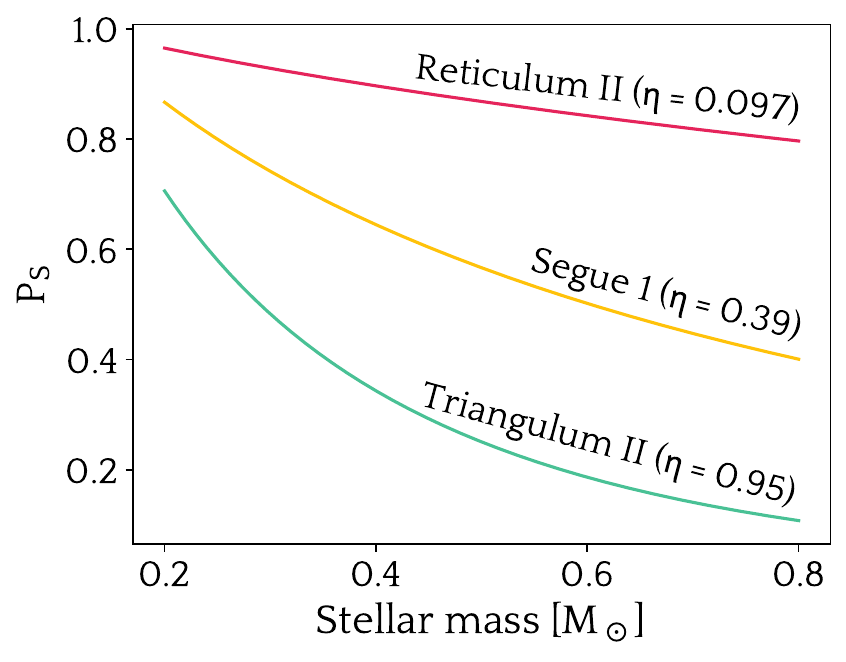}
\caption{\label{fig:surviprob} Probability $P_S$ of star survival against destruction by PBHs, as a function of the stellar mass. We consider three UFD galaxies: Reticulum II (upper curve), Segue 1 (middle curve), and Triangulum II (lower curve) and assume $f_\text{PBH}=1$.}
\end{figure}

In Fig.~\ref{fig:surviprob}, we show the probability of stars surviving destruction by PBHs (Eq.~\eqref{eq:surviprob}) in the three galaxies of our sample, assuming that PBHs constitute all the DM. This plot indicates that, for $f_\text{PBH}=1$, the star-destruction effect of PBHs is negligible in Reticulum II $(\eta=0.097)$, moderate in Segue 1 $(\eta=0.39)$, and strong in Triangulum II $(\eta=0.95)$. According to this observation, we use the galaxies Segue 1 and Triangulum II to test the PBH abundance, while Reticulum II is used as a 'control sample' as explained in detail in Sec.~\ref{sec:results}. 

\subsection{HST photometric observations}
We used photometric observations of the central regions of the UFD galaxies Reticulum II, Segue 1, and Triangulum II from the HST-ACS Treasury Program GO-14734 (PI: N. Kallivayalil). Here, we summarise some relevant details of the data processing. A more in-depth description can be found in \cite{Filion_2024} and \cite{Richstein_2024}.

The imaging data for each galaxy were taken in the F814W and F606W filters of the ACS instrument with the Wide Field Channel. The exposures were then processed to only retain the starlike sources, removing most of the other light sources, which can be attributed to contaminants such as background galaxies or cosmic rays. Note that these remaining sources may be not only single stars, but also unresolved binary systems. Among the remaining sources in the field, it is important to identify the stars, which are members of the targeted galaxy. To do this, we relied on the use of the Dartmouth Stellar Evolution Database \citep{Dotter}, which provides isochrone models for the position of stars of various masses on the colour-magnitude diagram (CMD) for a given age and metallicity.

\begin{figure}[h!]
\includegraphics[width=1\columnwidth]{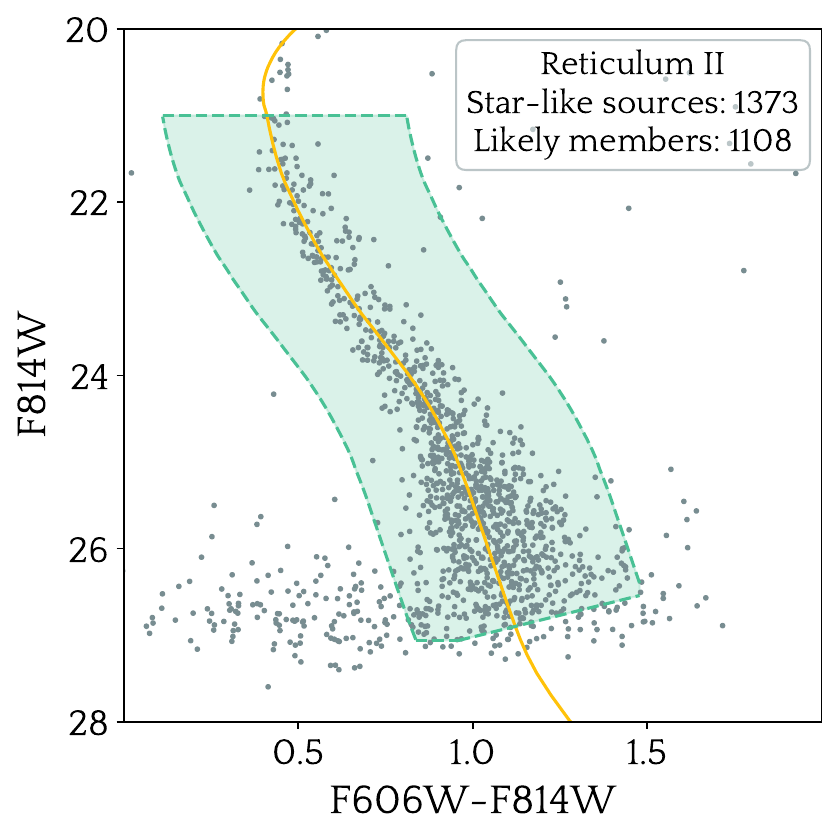}
\caption{\label{fig:CMD} CMD of Reticulum II. The solid line represents the Dartmouth isochrone for a population of $13$~Gyr old stars with metallicity [Fe/H]$=-2.5$, corrected for distance and extinction. The points correspond to the observed star-like sources, while the coloured region encompasses the stars that are likely members of the galaxy and will be used to study its stellar mass function.}
\end{figure}

We generated a solar-scaled isochrone, i.e. with solar ratios of all chemical elements other than iron, for a $13$~Gyr old stellar population with a metallicity of [Fe/H]$=-2.5$, the lowest metallicity available in the Dartmouth Database and the closest to the mean metallicities measured in these galaxies (see Table 2 in \citealt{Filion_2024}). We then converted this isochrone, originally in absolute magnitude, to apparent magnitude by correcting for the distance and reddening based on the data from Table 1 of \cite{Filion_2024}. For the latter, we used the Spanish Virtual Observatory (SVO) filter-profile service\footnote{Available at \textcolor{blue}{\href{http://svo2.cab.inta-csic.es/theory/fps/}{http://svo2.cab.inta-csic.es/theory/fps/}}.} to convert the V-band extinction values $A_V$ into extinction values for the F814W and F606W pass bands. The shifted isochrone is depicted in yellow on the CMD in Fig. \ref{fig:CMD}.

Real sources deviate from this theoretical model due to various effects and are consequently spread around the isochrone curve. Following \cite{Filion_2024}, we therefore considered as likely member stars those sources that are bluer than the isochrone by less than 0.3 mag in colour (i.e. on the left side of the isochrone) or redder by less than 0.4 mag (i.e. on the right side of the isochrone). The larger value on the red side was chosen to account for unresolved binary stars, which generate points that are redder than single stars with the same luminosity.

Furthermore, the effect of data completeness had to be accounted for. \cite{Richstein_2024} performed artificial star tests to estimate the $50\%$ completeness limits of the UFD observations used in this work. To minimise uncertainties in star counts, we restricted the likely member sources to those with magnitudes below (i.e. brighter than) these limits, denoted by $\text{F814W}_\text{max}$ and $\text{F606W}_\text{max}$. Additionally, we restricted the likely members sample to stars that are expected to still be on the main sequence, according to the Dartmouth model, which corresponds to selecting stars with magnitudes above $\text{F814W}_\text{min}$. The values of these limits can be found in the Table 1 of \cite{Filion_2024}.

We obtained samples containing 1108, 559, and 608 likely member sources for Reticulum II, Segue 1, and Triangulum II, respectively\footnote{The difference between these numbers and those in \cite{Filion_2024} is due to roundings and updates of the values on the SVO website.}. As an example, we display in Fig. \ref{fig:CMD} the CMD of all the star-like sources of Reticulum II, alongside the theoretical isochrone model that is used to identify the likely members stars that lie within the coloured region on the plot. The CMDs of the other two galaxies can be found in Appendix \ref{app:CMDs}. The resulting catalogues of likely members are used in the following section to study the stellar mass function and the possible presence of asteroid-mass PBHs in these galaxies.

\section{Bayesian inference procedure}
\label{sec:Bayesian}
\subsection{Synthetic stellar populations}
\label{sec:synthetic_pop}
The observational data takes the form of CMDs, with each source (which can be either a single star or an unresolved binary system) providing a magnitude value in both the F606W and F814W pass bands. However, the model present-day mass function from Sec. \ref{sec:IMF} describes the distribution of stars as a function of their masses, not their magnitudes. Unfortunately, it is impossible to accurately recover stellar masses from the observed magnitudes, as one would need to know the metallicity of each individual star, identify which sources are in fact unresolved binaries, and know what the photometric error associated with each measured magnitude is. Instead, we used the 'forward modelling' strategy, as described in \cite{Filion_2022}, which allowed us to compare the real observed magnitude values with synthetic ones constructed from a model IMF.

To produce these mock observations, we first generated masses according to the present-day mass function as given in Eq.~\eqref{eq:PDMF} above. This mass function takes into account the effect of PBHs and depends on three parameters: the two IMF parameters represented by $\theta$ of Sec.~\ref{sec:IMF_only}, and the PBH fraction $f_\text{PBH}$. A fourth parameter, the binary fraction\footnote{In this work, the binary fraction, $f_b$, is defined as the ratio of the number of actual stars that are members of binary systems to the total number of stars. However, another convention has sometimes been used in previous IMF analyses \citep[e.g.][]{Geha}, in which the binary fraction $f_g$ is defined as the fraction of sources in the sky that are, in fact, binaries rather than single stars. This definition can be recovered using $f_g=f_b/(2-f_b)$.} $f_b$, was also specified and used to randomly pair some of these masses together. We then drew a metallicity for each star from a truncated Gaussian distribution with mean, dispersion, and truncation points specified by the values in Table 2 of \cite{Filion_2024}. We ensured that both stars in a given binary system were assigned the same metallicity.

Using isochrone curves from the Dartmouth Stellar Evolution Database \citep{Dotter}, each mass and metallicity combination was converted into absolute F814W and F606W magnitudes by interpolating the solar-scaled $13$~Gyr old isochrones. For stars that were attributed a metallicity [Fe/H] $<-2.5$, we used the isochrone with the lowest available metallicity, i.e. [Fe/H]$=-2.5$. The total magnitude of a binary was obtained by summing the fluxes of both stars. Based on the distance and reddening measurements in Table 1 of \cite{Filion_2024}, we then converted the absolute magnitudes into apparent magnitudes. We modelled the uncertainties on these parameters by adding a random Gaussian error, identical for all stars of a given population realisation, with a width of $20\%$ on $A_V$ and $1$ kpc on the distance before converting the magnitudes. We also modelled the effects of photometric errors by generating a Gaussian noise for each source. The mean of this Gaussian noise is $0$, and its dispersion was obtained by fitting the magnitude-dependent errors measured in both bands for the real data. Lastly, to model the incompleteness of the real data, we applied rejection sampling to remove some stars using the completeness values derived in \cite{Richstein_2024} and following the functional form described in Appendix C2 of \cite{El-Badry_2017}. We also applied the same magnitude cuts as those used for the real data.

While stars were initially generated with masses in the range $[0.1,1]M_\odot$, the application of magnitude cuts restricts the low-mass end to approximately the completeness limit and the high-mass end to $\lesssim 0.8M_\odot$. However, low-mass stars ($\lesssim0.15M_\odot$) could still avoid the magnitude cuts and remain in the sample, provided they are members of binary systems. The expression \eqref{eq:nu}, originally derived for stars with masses $\in[0.2,0.8]M_\odot$, was, therefore, extrapolated to masses between $0.1$ and $0.2M_\odot$. It is also worth noting that the input binary fraction, $f_b$, may differ slightly from the final fraction observed in the mock dataset, as single and binary stars do not share the same magnitude distribution and, thus, have different probabilities of being excluded by the magnitude cuts.

\begin{figure}[h!]
\includegraphics[width=1\columnwidth]{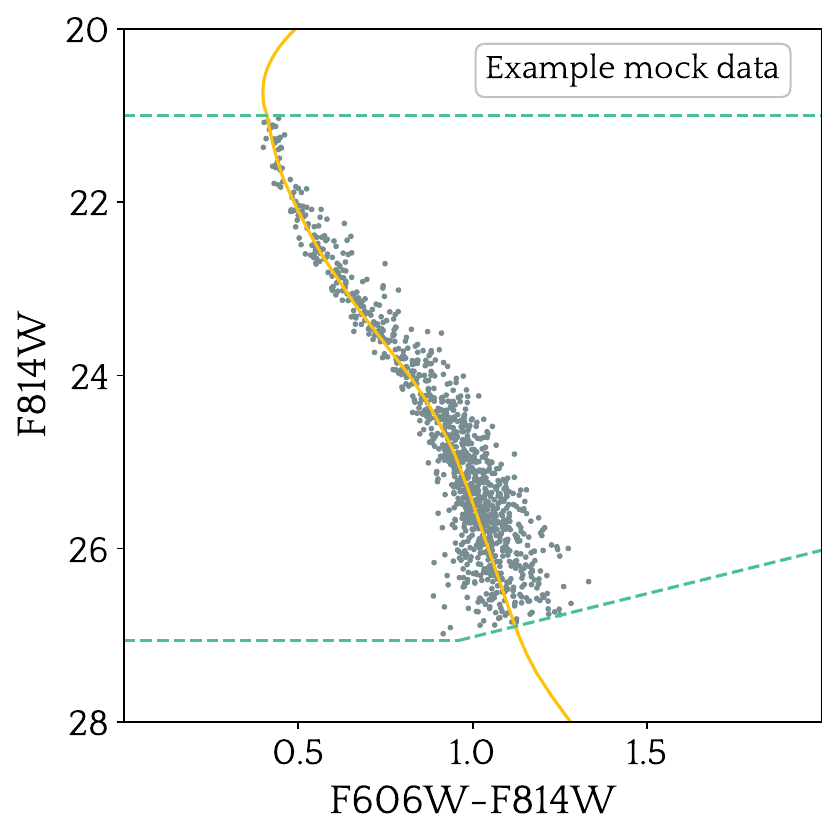}
\caption{\label{fig:CMD_mockdata} CMD of a mock dataset. The solid line represents the Dartmouth isochrone for a population of $13$~Gyr old stars with metallicity [Fe/H]$=-2.5$, corrected for distance and extinction. The points correspond to the generated star-like sources. This mock population was generated with the magnitude cuts (as shown by dashed lines) and physical parameters of Reticulum II, a Kroupa IMF, a binary fraction of 0.5, and under the assumption of no PBH effect $(f_\text{PBH}=0)$.}
\end{figure}

Thus, the generated synthetic stellar populations depend on four parameters: the two IMF parameters $\theta$, the binary fraction $f_b$, and the fraction of PBHs $f_\text{PBH}$. We ensured that the mock datasets contained the same number of sources as the real observations they were supposed to mimic, in order to then create a CMD similar to that in Fig.~\ref{fig:CMD}, but for the mock datasets. An example is displayed in Fig.~\ref{fig:CMD_mockdata}.

\subsection{Approximate Bayesian computation}
To compare the real CMD of a given galaxy (Fig.~\ref{fig:CMD}) with the synthetic CMDs similar to the one shown in Fig.~\ref{fig:CMD_mockdata}, we divided the unbinned CMDs into square bins of the size $0.2$ mag. Both the observed and synthetic datasets were then represented as a list of $\mathcal{O}(100)$ values of (normalised) source counts within bins. We chose the Jensen-Shannon distance between the mock and real lists as our distance metric. It was checked in \cite{Filion_2022} and \cite{Filion_2024} that the use of other bin sizes or distance metric had no major impact on the results.

With this setup, we obtain, for each realisation of the mock dataset with given input parameters $\{\theta,f_b,f_\text{PBH}\}$, a number -- the Jensen-Shannon distance -- indicating how close this mock dataset is to the real data. Because of the many random variables involved in the generation of mock datasets, this number fluctuates even at fixed model parameters. This same randomness, coupled with the complex dependencies of stellar mass and metallicity on magnitude, makes the task of defining a likelihood function for the model populations in magnitude space difficult. We thus performed a likelihood-free approximate Bayesian computation (ABC), which aims to find the model parameters that minimise the average distance between a large sample of mock datasets and the real data.

The workflow of the ABC using a sequential Monte Carlo scheme (ABCSMC) as implemented within pyABC \citep{schaelte2022pyabc} is the following. We chose a prior for each parameter (see Sec.~\ref{sec:results}) and sample from these priors, generating  $\mathcal{N}=1000$ parameter values and corresponding star populations as described in Sec.~\ref{sec:synthetic_pop}. For each, we calculated the associated Jensen-Shannon distance to the real data, thus obtaining an ensemble of 'particles', where each particle represents a point in the parameter space and its corresponding distance to the data. The 90\% quantile of smallest distances defines the acceptance threshold; 
particles above the threshold (10\% of the original number) were discarded. The remaining set was used to generate a new ensemble of $\mathcal{N}$ particles via importance sampling method, from which point the next iteration starts. We assumed that the algorithm has reached convergence once three subsequent iterations have threshold values differing by less than $0.5\%$. We then ran five additional steps, followed by one final step with an ensemble of $10\times\mathcal{N}$ particles, whose distribution in the parameter space is the posterior distribution we are looking for. In the analysis of Triangulum II (see Sec.~\ref{sec:SegueTri}), we instead used $250\times\mathcal{N}$ particles in the final step.

To ensure that the algorithm worked properly, we generated several synthetic datasets and treated them as real data, aiming to recover the input parameters using the ABCSMC scheme. We tested populations ranging from 500 to 3000 stars, with various binary fractions, both the BPL and LN stellar mass function forms, and scenarios with and without the PBH effect. In all the tests the parameters $\{\theta,f_b,f_\text{PBH}\}$ used to generate the synthetic data were consistently recovered within their statistical uncertainties. 

\section{Results and constraints on PBHs}
\label{sec:results}
\subsection{Reticulum II: The control galaxy}
\label{sec:ret2_control}

We first turn to Reticulum II, a galaxy with a low merit factor for which the effect of PBHs is expected to be negligible. Therefore, the present-day stellar mass function of this galaxy should be approximately equal to its IMF. We performed the ABCSMC analysis described in Sec. \ref{sec:Bayesian} for the two IMF parameters, the binary fraction, and the PBH fraction, with uniform priors $\alpha_{1,2}\in[-4.5,-0.05]$, $M_c\in[0.05,1]$, $\sigma_\text{LN}\in[0.25,1]$, $f_b\in[0,1]$, and $f_\text{PBH}\in[0,3]$. These ranges cover a broad spectrum around the typical values of these parameters reported in previous studies \citep{Geha,Gennaro_2018a,Gennaro_2018b,Filion_2022,Filion_2024}.

\begin{figure}[h!]
\includegraphics[width=1\columnwidth]{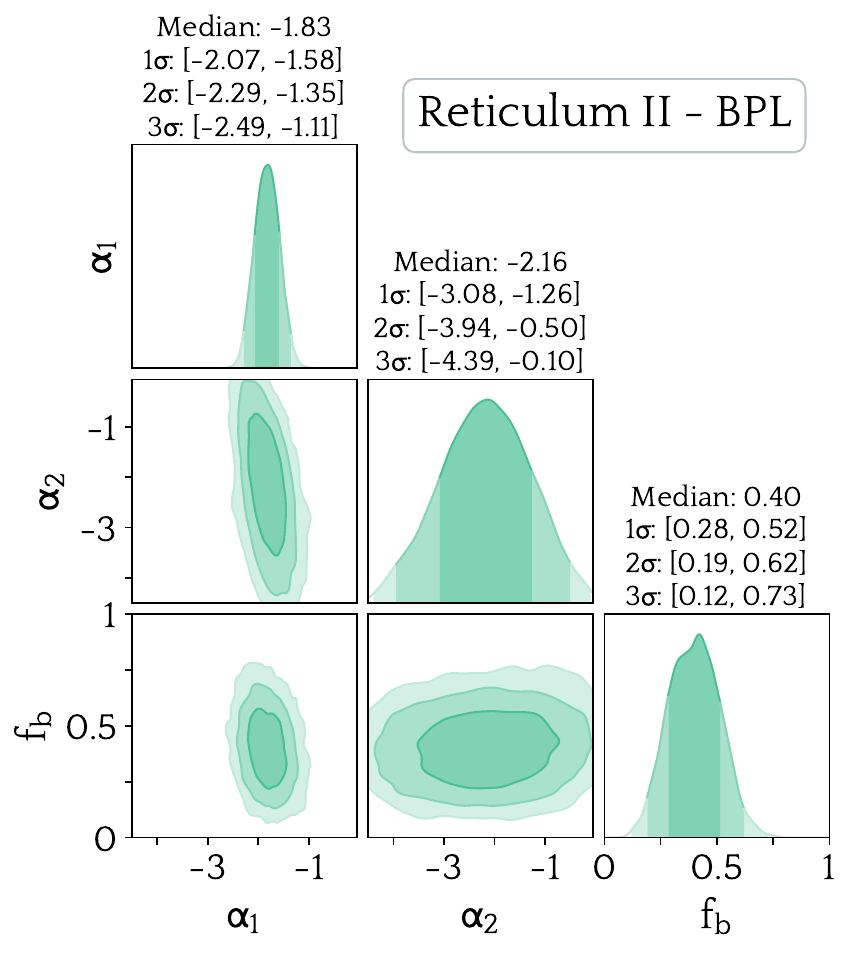}
\includegraphics[width=1\columnwidth]{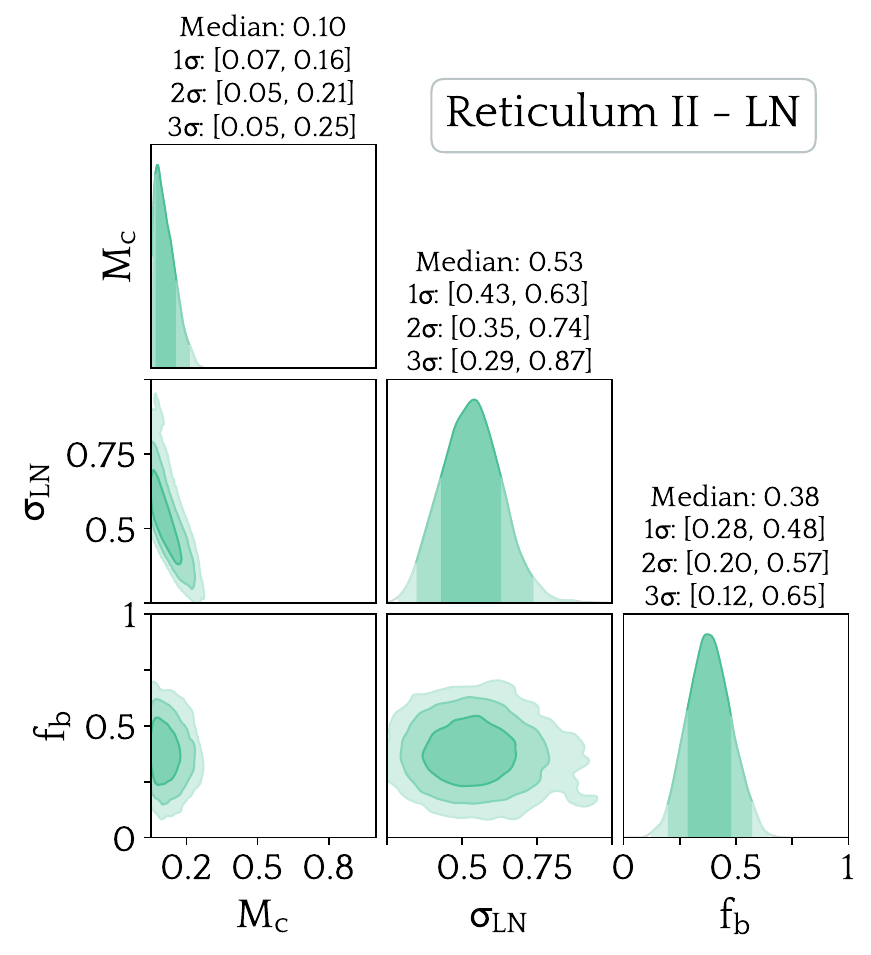}
\caption{\label{fig:ret2} Corner plots for the galaxy Reticulum II for two different IMFs: the BPL (upper panel) and LN (lower panel) distributions. The confidence levels, centred on the median, are represented by $1\sigma$, $2\sigma$, and $3\sigma$ intervals, with colours ranging from darker to lighter shades. The results are independent of $f_\text{PBH}$ because of the low merit factor of Reticulum II.}
\end{figure}

The posterior distributions of the IMF parameters for both the BPL and the LN mass functions are shown in Fig. \ref{fig:ret2}. The posterior distribution of $f_\text{PBH}$ was found to be roughly uniform across its allowed range and is therefore not displayed on the plot. This is the expected behaviour for this parameter since it does not influence the model due to the low merit factor of Reticulum II. We also verified that performing the ABCSMC analysis with fixed $f_\text{PBH}=0$ yielded the same posterior distributions for the other parameters, thus confirming that the PBH impact on Reticulum II is negligible.

These results are consistent within $1\sigma$ with those obtained by \cite{Filion_2024}, with slightly tighter posteriors. These plots indicate that the most constrained parameters are the low-mass slope $\alpha_1$ for the BPL and the characteristic mass $M_c$ for the LN IMF. This can be attributed to the fact that most stars in the samples have low masses, providing more data points to constrain these parameters, which are predominantly influenced by the low-mass data. 

Another notable observation is the near absence of correlation between the IMF parameters and the binary fraction. While the IMF parameters primarily determine the distribution of points in luminosity (i.e. along the vertical axis of the CMD), the binary fraction influences how spread the data points are in colour (i.e. along the horizontal axis of the CMD), as binary stars appear redder than single stars with the same total luminosity. 

Lastly, modifying the isochrone model, the extinction, or the distance of the galaxy mainly results in theoretical curves that are shifted along the horizontal axis relative to those used here. We thus expect that among model parameters the binary fraction is most sensitive to the stellar population modelling. The other parameters should be fairly robust against systematic errors arising from the choice of isochrone. This concerns equally the PBH fraction whose main effect on the CMD is also along the vertical axis. 

\subsection{Segue 1 and Triangulum II: The test galaxies}
\label{sec:SegueTri}

In the analysis of Reticulum II in Sec.~\ref{sec:ret2_control}, we used broad uniform priors for all parameters. But, in general, there is no compelling reason to favour a flat prior over another function. Previous mass function analyses have explored other possibilities, such as a logarithmic prior \citep{Gennaro_2018a}.  

For the particular question we address in this paper, however, there is a well-motivated choice. The three galaxies in our sample share several key properties: they have comparable ages, metallicities, and velocity dispersions, differing primarily in their inferred DM densities (see Table~\ref{tab:data1}). While Reticulum II has multiple r-process enriched stars \citep{Ji_2016} and is likely a satellite of the Large Magellanic Cloud \citep{Patel_2020} as opposed to the other two UFDs, which are Milky Way satellites, all three galaxies share comparable star formation histories \citep{Sacchi_2021,Simon_2023}. Therefore, there is no reason to expect substantial variations in the mass function among these three systems. This assumption can be tested by applying the analysis of Sec.~\ref{sec:ret2_control}, with $f_\text{PBH}=0$, to the other two galaxies. The results are presented in Appendix \ref{app:full_pdfs_noPBHs}. The inferred values of the IMF parameters are consistent within their statistical uncertainties among the three galaxies.

In contrast to other parameters, the merit factors (and, consequently, the expected effect of PBHs) in these galaxies differ significantly, being negligible in Reticulum II and substantial in Segue 1 and Triangulum II. If PBH effects were sizeable, it would be difficult to explain the good agreement observed in the inferred IMF parameters among these three galaxies. Hence, constraints on the PBH fraction, $f_\text{PBH}$, can be derived. 
This suggests the following 'control sample' strategy: one may use the posterior distributions of the IMF parameters and binary fraction inferred from Reticulum II as priors when analysing Segue 1 and Triangulum II.

Using these 'control sample' priors for the three first parameters and uniform $f_\text{PBH}\in[0,3]$ -- which allows PBHs to make up to $300\%$ of the DM to conservatively avoid boundary effects on the tail of the posterior distribution -- we performed the ABCSMC analysis from Sec. \ref{sec:Bayesian} for both Segue 1 and Triangulum II, and the two parametrisations of the IMFs. The resulting posterior distributions are displayed in Appendix \ref{app:full_pdfs}. We show here in Fig. \ref{fig:tr2+sg1} only the marginalised distributions for our parameter of interest $f_\text{PBH}$.

\begin{figure}[h!]
\includegraphics[width=\columnwidth]{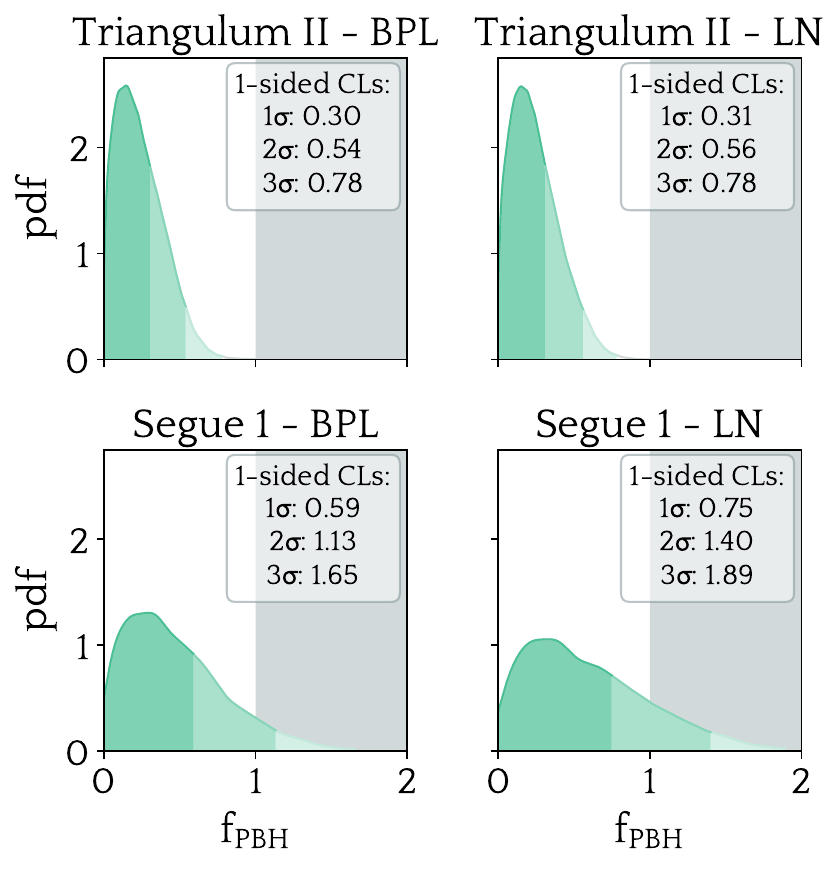}
\caption{\label{fig:tr2+sg1} Marginalised posterior probability density functions for the fraction of DM in PBHs,  $f_\text{PBH}$, for the galaxies Triangulum II (upper panels) and Segue 1 (lower panels), adopting the BPL (left panels) and LN (right panels) IMFs. One-sided $1\sigma$, $2\sigma$, and $3\sigma$ intervals are displayed with colours ranging from darker to lighter shades. The grey region corresponds to $f_\text{PBH}>1$.}
\end{figure}

Using one-sided exclusion levels, we find that for both forms of the stellar IMF, the galaxy Segue 1, for which the effect of PBHs is moderate, is not very constraining, as the value $f_\text{PBH}=1$ is excluded at the level of slightly less than $2\sigma$. In contrast, Triangulum II excludes PBHs as constituting more than $\sim 55\%$ ($\sim 78\%$) of the DM at a $2\sigma$ ($3\sigma$) confidence level. 
The value $f_\text{PBH}=1$ is excluded by Triangulum II at a $3.7\sigma$ ($4.1\sigma$) level for the BPL (LN) IMF. The results are overall quite similar for both IMFs, indicating that they are robust against the choice of the IMF law.

We also conducted the ABCSMC analysis using uniform priors for all parameters and found that $f_\text{PBH}\gtrsim 1.2$ was excluded at the $2\sigma$ level by Triangulum II in that case. This weaker exclusion can be attributed to the degeneracy between $f_\text{PBH}$ and $\alpha_1$ or $M_c$, depending on the mass function parameterisation. Uniform prior permits particularly high values of $\alpha_1\gtrsim-0.8$, which are not commonly observed. Such a shallow slope in the BPL IMF would result in an overabundance of high-mass stars in the synthetic populations. Consequently, a high value of $f_\text{PBH}$ would be required to destroy these excess high-mass stars and reproduce the observed present-day mass function. The same effect can be found for unrealistically high values of $M_c\gtrsim 0.6$, which also overproduce high-mass stars. To illustrate the degeneracy between $f_\text{PBH}$ and $\alpha_1$ or $M_c$ when uniform priors are used, we show in Appendix \ref{app:f_degeneracy} the corner plots for these parameters, for both Triangulum II and Segue 1. Again, we emphasise that the uniform priors are less motivated than the 'control sample' priors we used in our analysis, which exclude these unrealistically high values of $\alpha_1$ and $M_c$.

Lastly, we also performed an analysis using priors from a larger control sample. These new priors were obtained by combining the posterior distributions resulting from the analysis of Reticulum II and two other galaxies for which the data are available, and where the impact of PBHs is expected to be negligible, namely Ursa Major II and Boötes I. Although these galaxies are not as similar to our test galaxies as is Reticulum II, the priors obtained this way still exclude excessively high values of $\alpha_1$ and $M_c$, which resulted in $f_\text{PBH}=1$ being excluded at more than $3\sigma$ by Triangulum II. This demonstrates that our results are not solely due to using Reticulum II as a source of priors. 

\begin{figure}[h!]
\includegraphics[width=1\columnwidth]{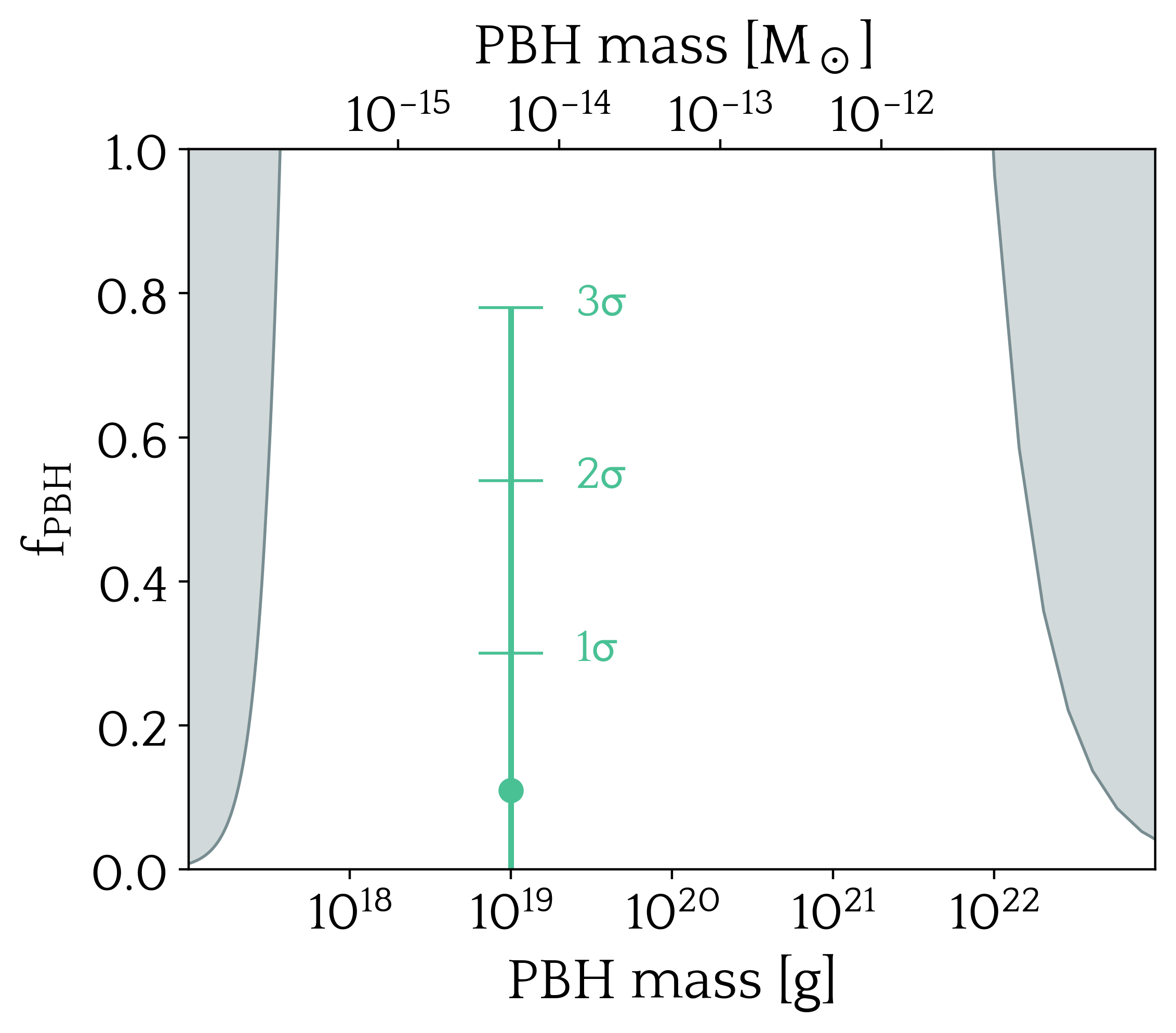}
\caption{\label{fig:constraints} Constraints on $10^{19}$g asteroid-mass PBHs from this work. The dot corresponds to the maximum of the posterior distribution of $f_\text{PBH}$, whose value is consistent with $f_\text{PBH}=0$. To the left and the right, we show the existing constraints on PBHs arising from evaporation and microlensing, plotted using the data compiled in the PBH bounds repository (\textcolor{blue}{\href{https://github.com/bradkav/PBHbounds}{https://github.com/bradkav/PBHbounds}}).}
\end{figure}

In Fig.~\ref{fig:constraints}, we present the constraints from Triangulum II (with the BPL mass function) on asteroid-mass PBHs around $10^{19}$g. We note again that, due to the uncertain effect of binary stars on PBH capture for masses above this range, it is not clear at the moment how to extend the constraints to higher masses. A more detailed study of binary systems may enable us to extend these constraints up to $10^{21}$~g in the future.

\section{Conclusion}
\label{sec:conclusion}

In this work, we derived observationally based constraints in the previously open asteroid-mass window of PBHs, hence, showing that PBHs with masses around $10^{19}$g cannot constitute all of the DM. We relied on the impact of these PBHs on the stellar mass function of UFDs and made use of photometric observations from the HST. 

\begin{figure}[h!]
\includegraphics[width=1\columnwidth]{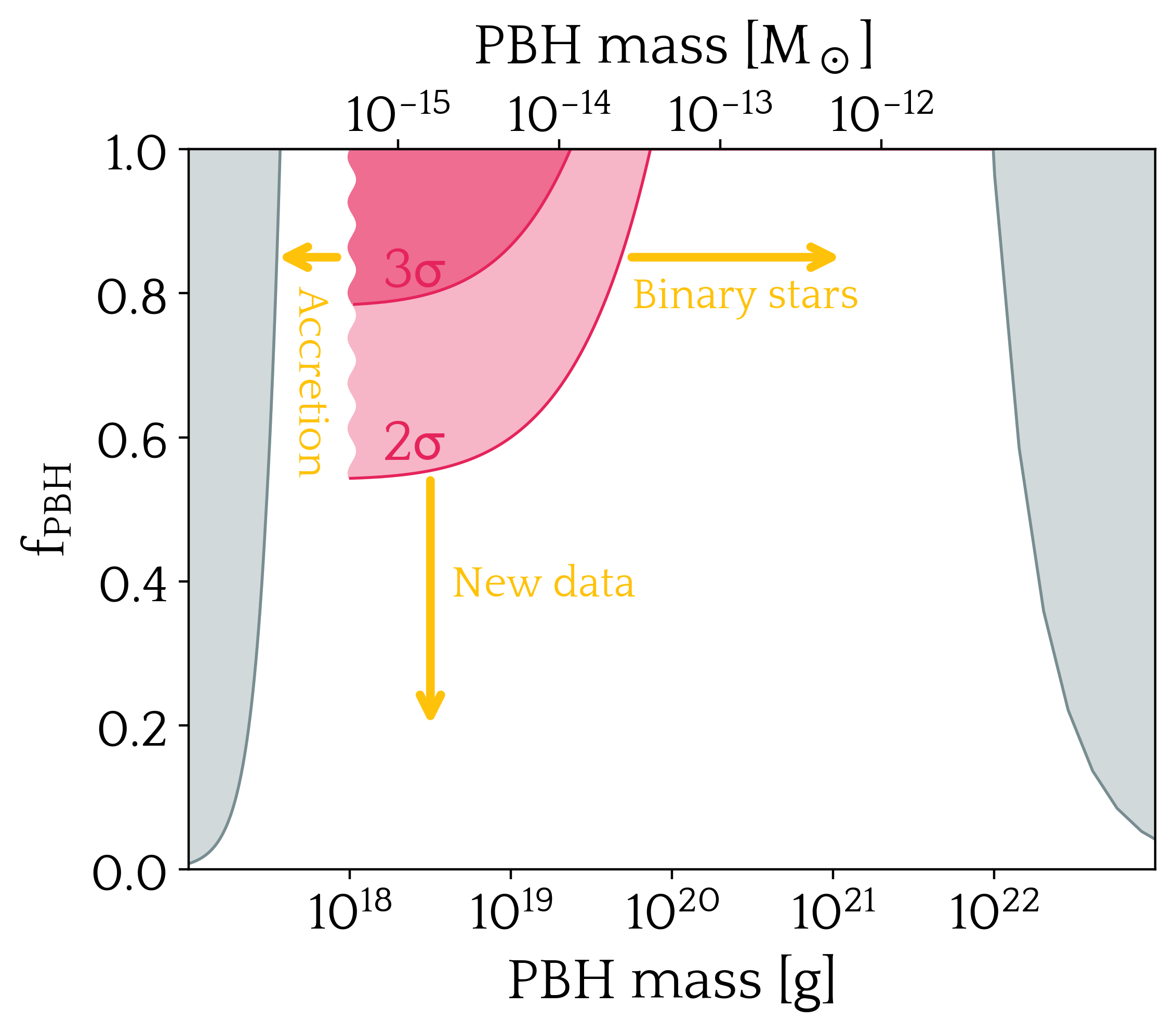}
\caption{\label{fig:constraints_extended} Extended constraints on asteroid-mass PBHs between $\sim 10^{18}$ and $\sim 10^{20}$ g. The red-coloured regions correspond to the $2\sigma$ (lighter) and $3\sigma$ (darker) excluded values of $f_\text{PBH}$. The arrows indicate the directions in which the constraints may be improved in the future. To the left and the right, we show the existing constraints on PBHs arising from evaporation and microlensing.}
\end{figure}

Assuming that the mean number of captured PBHs (Eq.~\ref{eq:nu}) is unchanged for PBH masses $\lesssim 10^{19}$ g, which is valid up to a $\sim30\%$ error \citep{Esser1,Tinyakov2024}, and using conservatively the bound derived in Appendix \ref{app:binary} for unperturbed PBHs in binary systems, the constraint from Sec.~\ref{sec:SegueTri} can be extended to PBH masses between $\sim 10^{18}$ and $\sim 10^{20}$ g, as displayed in Fig.~\ref{fig:constraints_extended}. While currently the constraints are marginal and cover a limited range of masses, future theoretical and observational advancements are expected to significantly improve these results, as indicated by the arrows in the figure.

In order to extend the constraints to lower masses, the accretion of stars by PBHs located in their cores must be studied more carefully. In \cite{Esser1}, the \cite{Bondi_1952} model was used to estimate the accretion time. It was found that the star destruction time scales inversely with the PBH mass, approaching $\sim 10$~Gyr for masses $\lesssim 10^{18}$~g. In this mass range, factors of order 1 become significant, so a more thorough analysis -- taking into account the radiation feedback within the accretion sphere and the rotation of the star -- is required for a more accurate estimate of the destruction time. Some authors have already partially studied the subject (see e.g. \cite{Markovic_1995}). 

To understand the behaviour for masses above $\gtrsim 10^{19}$ g, the destruction probability of stars in binary systems by PBHs must be calculated. For now, we calculated in Appendix \ref{app:binary} the number of captured PBHs that are unaffected by the presence of a companion star, and conservatively assumed that all other PBHs are not captured. However, this is most likely not true, and a dedicated three-body study of the binary star-PBH system is needed to understand the impact of binaries on the capture rate.

On the experimental side, there are multiple opportunities for advancement that may yield more stringent constraints in the near future.
First, the PBH capture probability grows with decreasing velocity dispersion. New spectroscopic measurements of UFDs will improve our determination of their velocity dispersions, specifically for Triangulum II, where only an upper limit on $\sigma_\text{los}$ exists at present. Using the actual value instead of the upper limit, as was done in this analysis, will definitely improve the constraints. Additionally, measuring velocity dispersions in other local dwarf galaxies may reveal new candidates where the PBH effect could be significant, or enlarge the number of galaxies which may be used as a 'control sample'. Several large surveys, such as those using the Prime Focus Spectrograph \citep{PFS} and the Dark Energy Spectroscopic Instrument \citep{DESI}, are expected to provide a substantial amount of new spectroscopic data in the coming years. 

Another avenue for improvement lies in photometry: while the HST already enables very deep photometry, resolving stars with masses as low as $\sim 0.2 M_\odot$ in some dwarf galaxies, the James Webb Space Telescope can resolve even fainter stars \citep{Weisz_2023}. This capability would increase the number of sources available for IMF analysis, yielding a more precise determination of the latter and, consequently, more stringent constraints on the PBH fraction.

With increasingly numerous and precise spectroscopic and photometric measurements of dwarf galaxies, future studies will allow comparisons across galaxies with varying merit factors. This could facilitate investigations into possible correlations between the merit factor and the IMF parameters, thereby constraining PBHs by comparing a broad sample of galaxies rather than focusing on individual cases.

Another possibility for probing asteroid-mass PBHs could arise from their impact, through the star destruction process, on the chemical evolution of UFDs. Alternatively, direct observations of star destruction by PBHs may provide observable smoking-gun signatures \citep{Markovic_1995,Bellinger_2023,Caplan_2023}.  However, much theoretical work is still needed in this direction to characterise the signal that may arise from such events.

\section*{Data availability}
The photometric catalogues used in this article may be obtained from the Mikulski Archive for Space Telescopes (MAST) at doi:\textcolor{blue}{\href{https://doi.org/10.17909/b5gn-6e22}{10.17909/b5gn-6e22}} (Reticulum II, Triangulum II, Segue 1, Ursa Major II) and doi:\textcolor{blue}{\href{https://doi.org/10.17909/t9-jr7h-en65}{10.17909/t9-jr7h-en65}} (Boötes I).
\begin{acknowledgements}
NE is a FRIA grantee of the Fonds de la Recherche Scientifique-FNRS and a member of BLU-ULB, the interfaculty research group focusing on space research at ULB - Université libre de Bruxelles. SDR is co-funded by the European Union (MSCA EDUCADO, GA 101119830). Views and opinions expressed are, however, those of the author(s) only and do not necessarily reflect those of the European Union. Neither the European Union nor the granting authority can be held responsible for them. The HST data were observed as part of Treasury Program GO-14734 (PI Kallivayalil). Support for this programme was provided by NASA through grants from the Space Telescope Science Institute,
which is operated by the Association of Universities for Research in Astronomy, Incorporated, under NASA contract NAS5-26555. PT is supported in part by the Institut Interuniversitaire des Sciences Nucl\'{e}aires (IISN) Grant No. 4.4503.15. RFGW is grateful for support through the generosity of Eric and Wendy Schmidt, by recommendation of the Schmidt Futures program. Computational resources have been provided by the Consortium des Equipements de Calcul
Intensif (CECI), funded by the Fonds de la Recherche Scientifique
de Belgique (F.R.S.-FNRS) under Grant No. 2.5020.11 and by the Walloon Region.
\end{acknowledgements}

\bibliographystyle{aa}
\bibliography{bibli}

\begin{appendix}

\begin{figure*}[h!]
\section{CMDs for Triangulum II and Segue 1}
\label{app:CMDs}
\includegraphics[width=1\columnwidth]{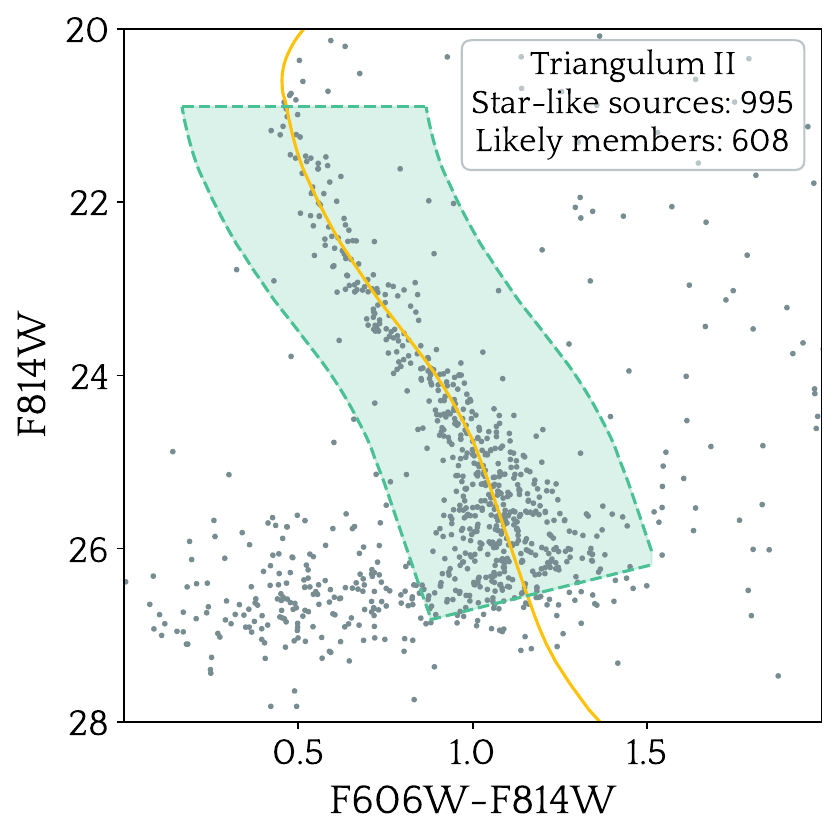}
\includegraphics[width=1\columnwidth]{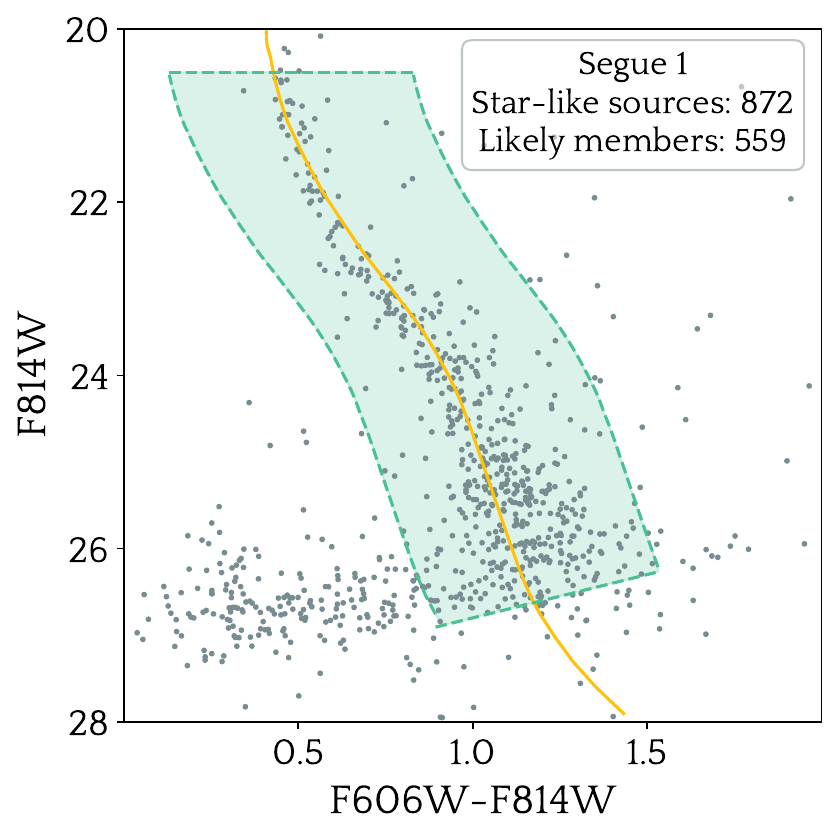}
\caption{\label{fig:CMD_sg1+tr2} CMDs of Triangulum II (left panel) and Segue 1 (right panel). The solid lines represent the Dartmouth isochrone for a population of $13$~Gyr old stars with metallicity [Fe/H]$=-2.5$, corrected for distance and extinction. The points correspond to the observed star-like sources, while the coloured regions encompass the stars that are likely members of the galaxies and will be used to study their stellar mass functions.}
\end{figure*}

\begin{figure*}[t]
\section{Corner plots for Triangulum II and Segue 1 -- no PBHs}
\label{app:full_pdfs_noPBHs}
\vspace{1cm}
\includegraphics[width=1\columnwidth]{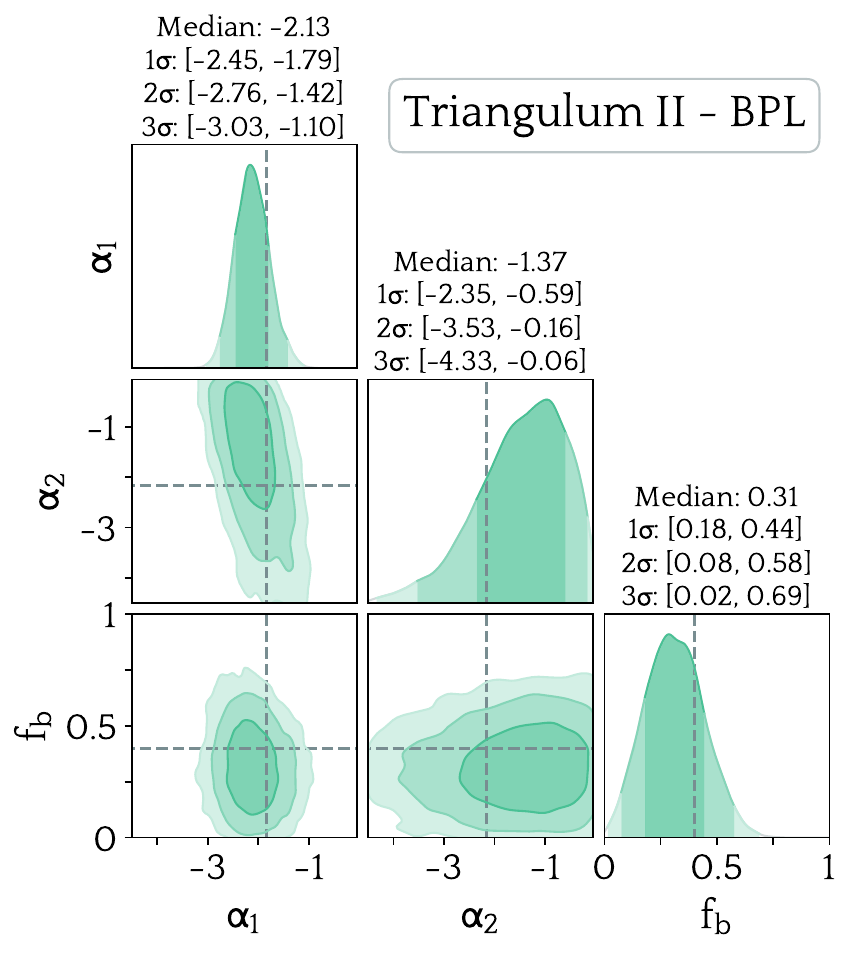}
\includegraphics[width=1\columnwidth]{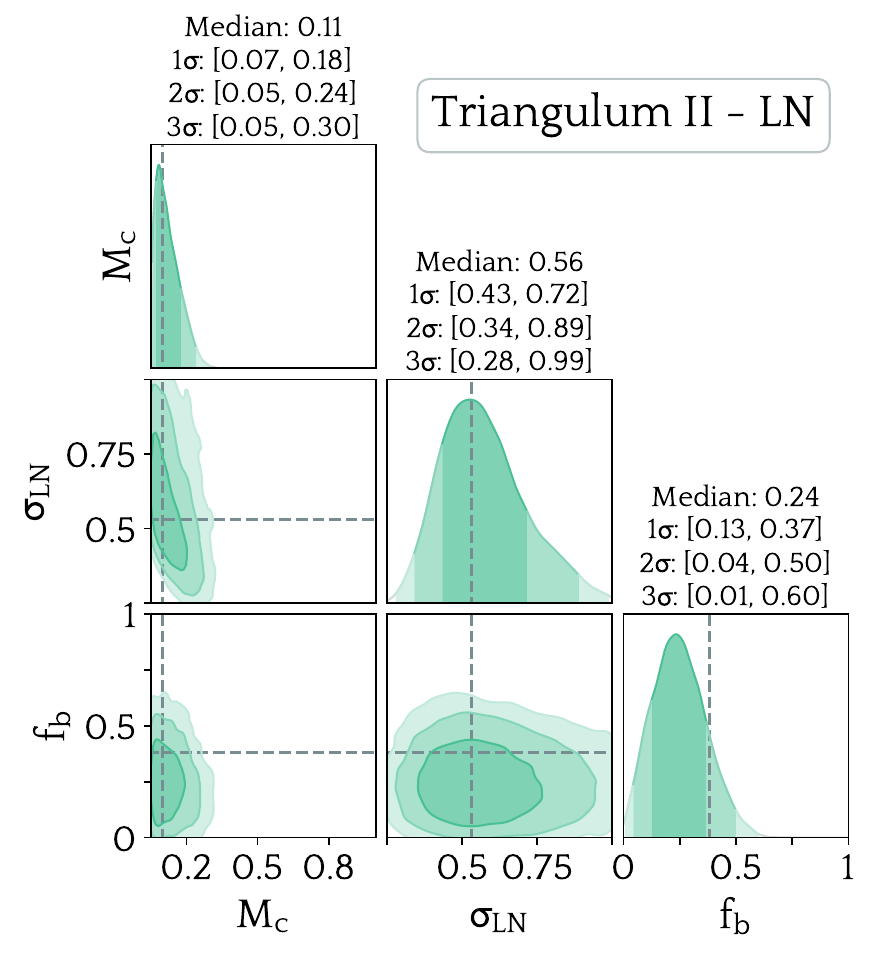}
\includegraphics[width=1\columnwidth]{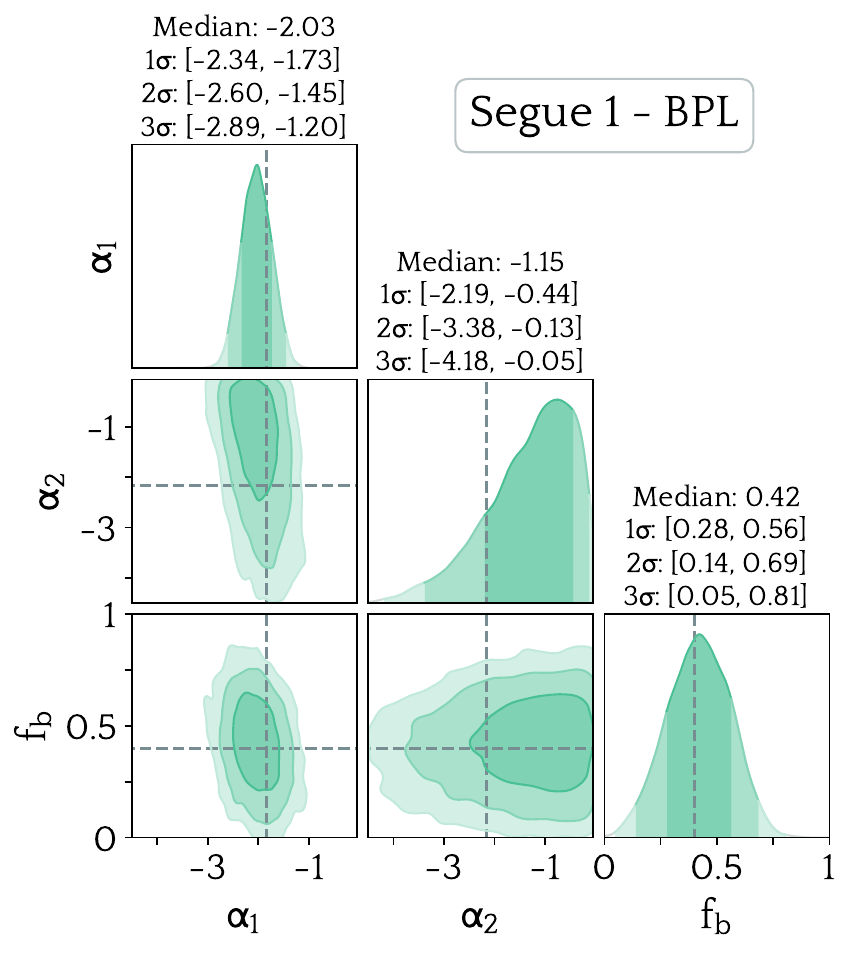}
\includegraphics[width=1\columnwidth]{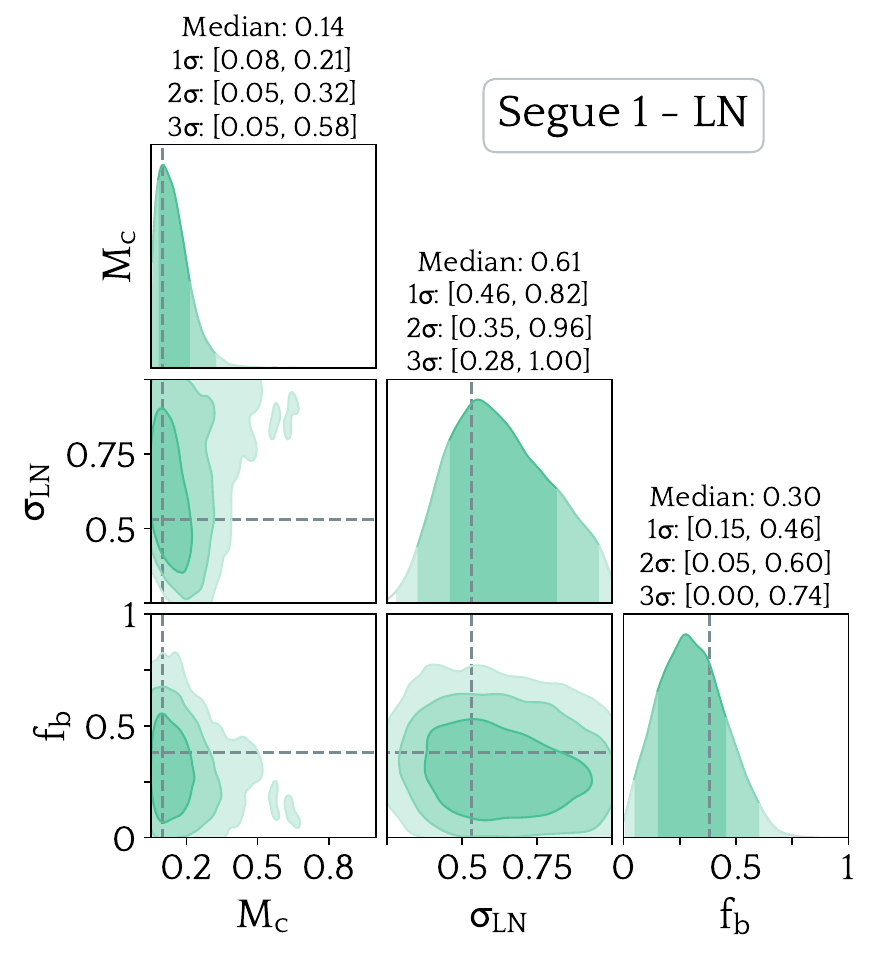}
\caption{\label{fig:tr2+sg1_BPL+LN_noPBH} Corner plots for the galaxies Triangulum II (upper panels) and Segue 1 (lower panels), with no PBH effect $(f_\text{PBH}=0)$, and for two different IMFs: the BPL (left panels) and LN (right panels) distributions. The confidence levels, centred on the median, are represented by $1\sigma$, $2\sigma$, and $3\sigma$ intervals, with colours ranging from darker to lighter shades. For easier comparison with Reticulum II (see Fig.~\ref{fig:ret2}), we show the median values of the parameters inferred for that galaxy as dashed lines.}
\end{figure*}

\begin{figure*}[t]
\section{Corner plots for Triangulum II and Segue 1 -- with PBHs}
\label{app:full_pdfs}
\vspace{1cm}
\includegraphics[width=1\columnwidth]{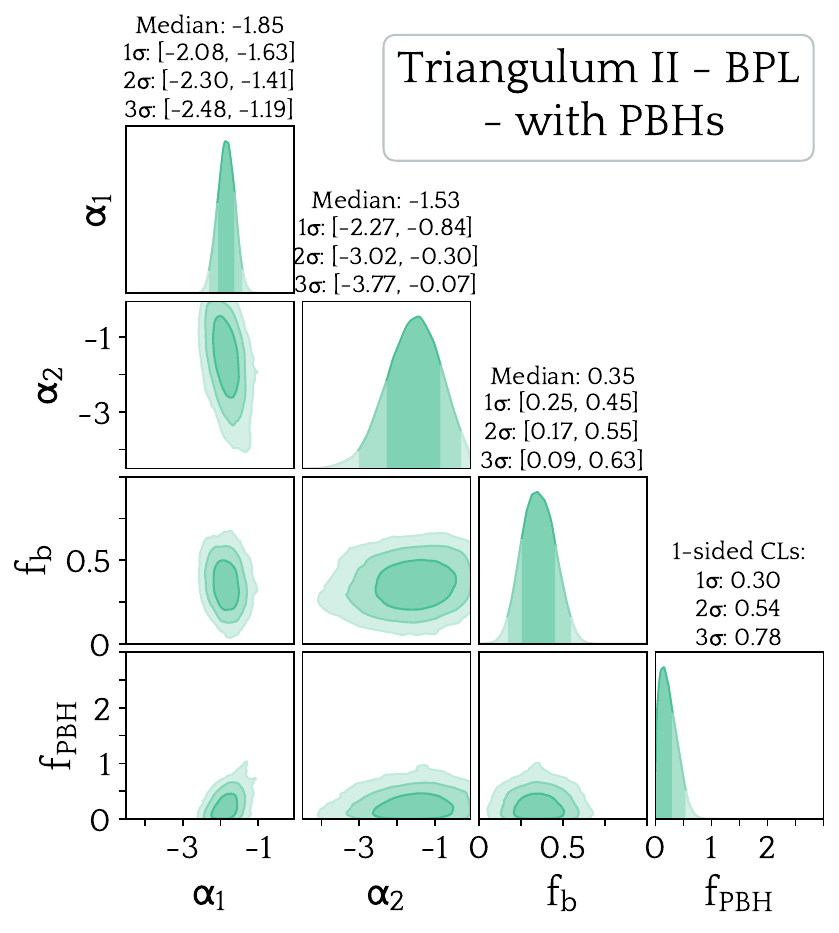}
\includegraphics[width=1\columnwidth]{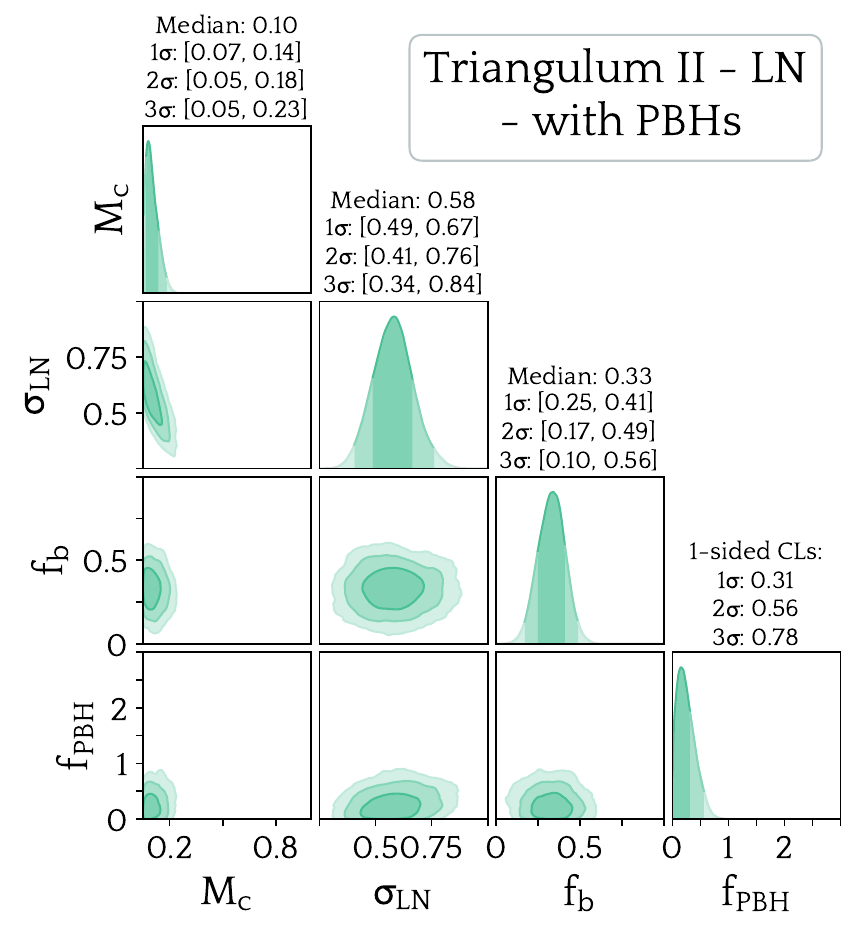}
\includegraphics[width=1\columnwidth]{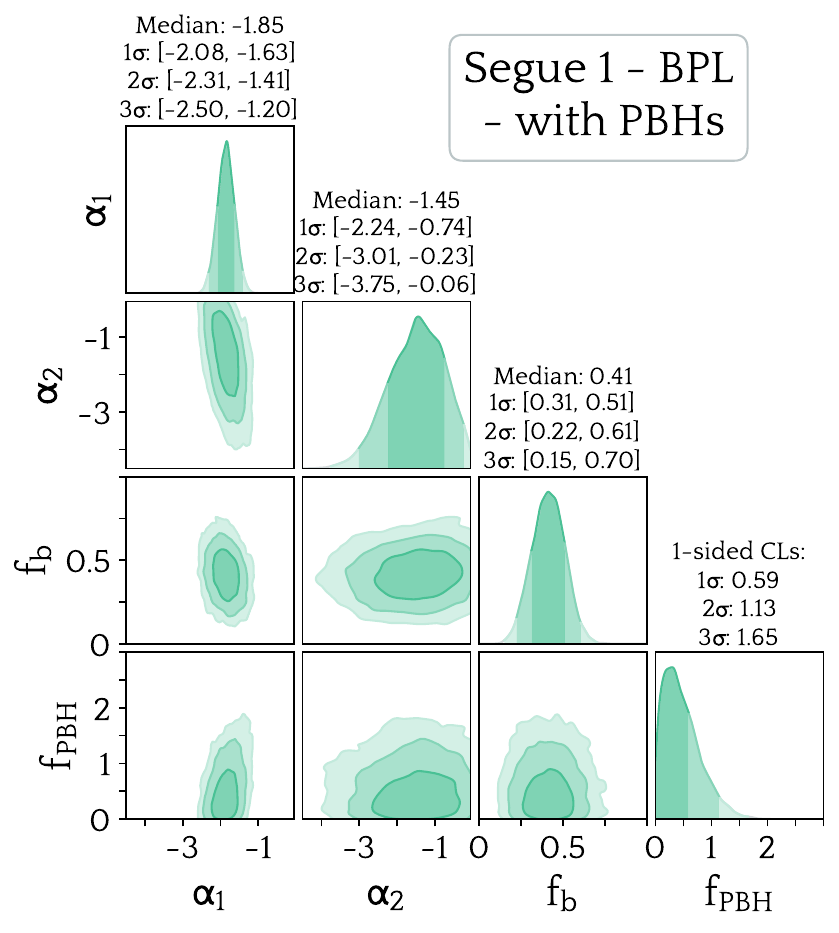}
\includegraphics[width=1\columnwidth]{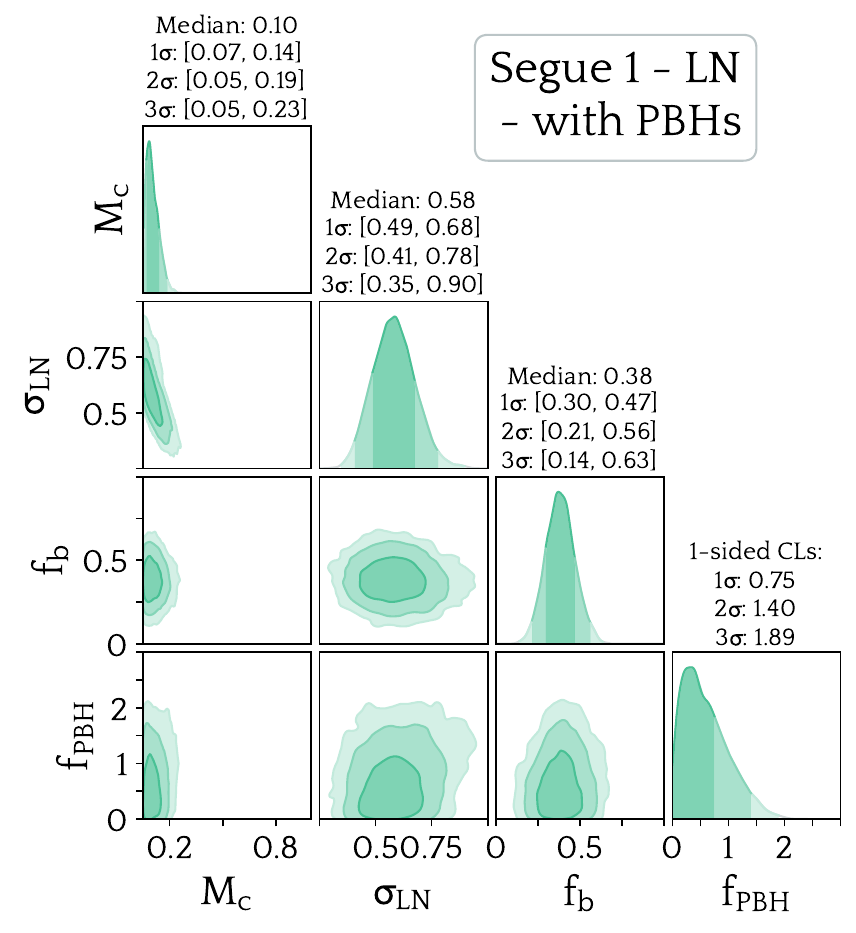}
\caption{\label{fig:tr2+sg1_BPL+LN} Corner plots for the galaxies Triangulum II (upper panels) and Segue 1 (lower panels), with PBH effect and for two different forms of the stellar IMF: the BPL (left panels) and LN (right panels) distributions. The confidence levels for all parameters are centred on the median, except for $f_\text{PBH}$ where the levels are one-sided. They are represented by $1\sigma$, $2\sigma$, and $3\sigma$ intervals, with colours ranging from darker to lighter shades. }
\end{figure*}

\FloatBarrier
\begin{figure}[h!]
\centering
\section{Plots of $f_\text{PBH}$ versus $\alpha_1$ or $M_c$ for Triangulum II and Segue 1 -- with PBHs -- uniform priors}
\label{app:f_degeneracy}
\includegraphics[width=1\columnwidth]{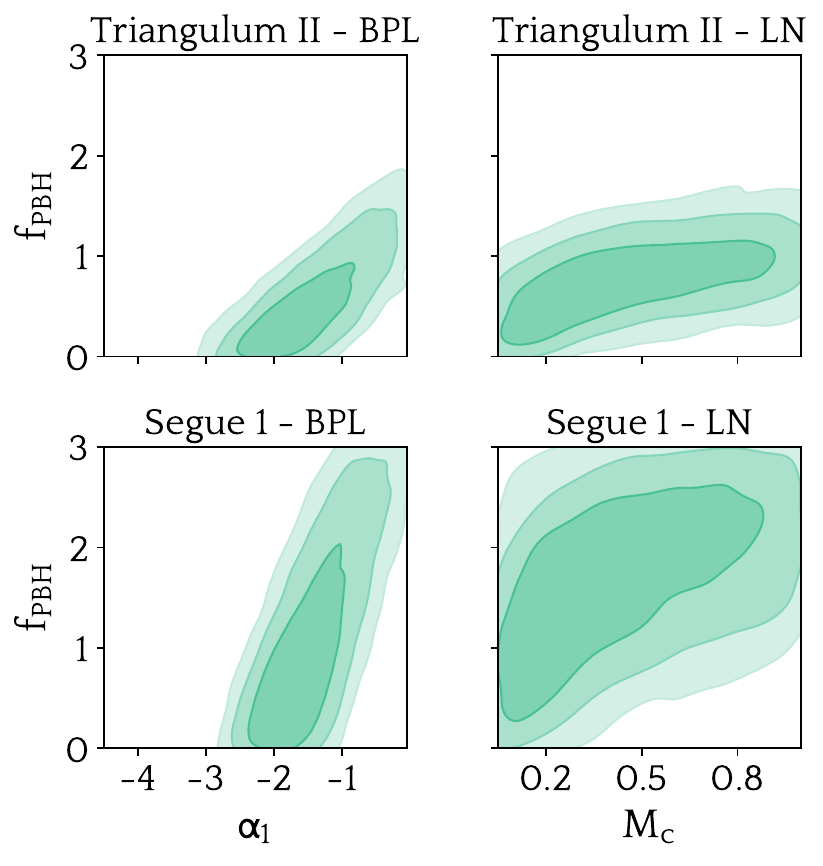}
\caption{\label{fig:f_degeneracy} Corner plots for the parameters $f_\text{PBH}$ vs $\alpha_1$ (BPL, left panels) and $f_\text{PBH}$ vs $M_c$ (LN, right panels), for the galaxies Triangulum II (upper panels) and Segue 1 (lower panels), with PBH effect and uniform priors. The confidence levels are represented by $1\sigma$, $2\sigma$, and $3\sigma$ intervals, with colours ranging from darker to lighter shades.}
\end{figure}

\section{PBHs and binary stars}
\label{app:binary}
\subsection{Analytical expression of \texorpdfstring{$\nu$}{TEXT} for single stars}
As was shown by numerical simulations in \cite{Esser1} and analytically in \cite{Oncins}, the number of PBHs orbiting around a newly formed star with Keplerian periastron $r_{\text{min}}$ and apastron $r_ {\text{max}}$ is given by \citep{Tinyakov2024}
\begin{equation}
\label{eq:NPBHcap}
    \frac{d\nu}{dr_{\text{min}} dr_{\text{max}}}=\frac{f_\text{PBH}\rho_{\text{DM}}}{m(\sqrt{2}\sigma_v)^3}(3\pi R_g)^{3/2}\frac{1}{\sqrt{2r_\text{max}}}
\end{equation}
with $m$ the PBH mass, $\sigma_v$ the velocity dispersion assumed to be the same as for stars, and $R_g=2GM$ the Schwarzschild radius of the star.

In the absence of close external perturbers such as a binary companion, the PBHs that will end up being captured by the star are those which \textit{(i)} have star-crossing orbits, i.e. $r_\text{min}<R$, and \textit{(ii)} lose enough energy due to the dynamical friction during each encounter with the star, so that their orbits become fully enclosed within the star in a time of order the age of stars in UFDs. This second condition can be expressed, assuming for simplicity that the star is uniform in density, and repeating the calculations of \cite{Esser1}, as follows:
\begin{equation}
    \label{eq:rT}
    r_{max}<r_T=\left(\frac{Tm\ln\Lambda}{\pi MR}\right)^2R_g\simeq142\left(\frac{m}{10^{20}\text{g}}\right)^2\text{AU},
\end{equation}
where $T=13$~Gyr is the time since star formation, $\ln\Lambda\simeq 30$ is the Coulomb logarithm and the numerical value on the right-hand side assumes $M=M_\odot$ and $R=R_\odot$. 

Integrating Eq. \eqref{eq:NPBHcap} over $r_\text{min}$ and $r_\text{max}$ with the conditions \textit{(i)} and \textit{(ii)}, we find the mean number, $\nu_\text{single}$, of PBHs that are captured by a single star in the absence of perturbers\footnote{The captured number, $\nu_\text{single}$, is the analytical equivalent of Eq.~\eqref{eq:nu}, except that it does not take into account the non-uniform density profile of the primary star, and the possible deviation of PBHs by distant perturbers (other than companion stars). We verified that the difference between the full numerical result used in the main text and this approximation is negligible.}. By combining Eqs.~\eqref{eq:NPBHcap} and \eqref{eq:rT}, we find that this number is independent of the PBH mass: while lighter black holes are more numerous at fixed $\rho_\text{DM}$, they experience less dynamical friction when colliding with the star and, thus, only those that begin with smaller orbits end up being captured. This also justifies why the survival probability formula \eqref{eq:surviprob} is valid in the range $m\in[10^{19},10^{21}]$g without depending on the PBH mass.

\subsection{Conservative bound on \texorpdfstring{$\nu$}{TEXT} for binary stars}
When a nearby perturber, such as a companion star, is present, a conservative condition for the PBH to be captured is that the angular momentum imparted to the PBH by the perturber remains sufficiently small, ensuring the PBH's periastron stays within the primary star's radius during two successive crossings. Assuming that the perturber is static over this time and located at a distance $D\gg r_\text{max}$ from the primary star, this condition can be written \citep{Esser1}
\begin{equation}
    r_\text{max}<r_D=\left(\alpha RD^6\right)^{1/7},
\end{equation}
where $\alpha$ is a $\mathcal{O}(1)$ calculable numerical coefficient depending on the direction of the perturber.

Extrapolating this formula to be valid for values of $D$ as small as $r_\text{max}$, and assuming conservatively that PBHs are not captured when $D<r_\text{max}$, one can obtain a lower bound on the number of PBHs that are captured by the primary star in the presence of a companion by integrating Eq. \eqref{eq:NPBHcap} for \textit{(i)} $r_\text{min}<R$ and \textit{(ii)} $r_\text{max}<r_C=\min(r_T,r_D,D)$. Of course, the result depends on the distance $D$ to the companion. Assuming that it is distributed according to a function $f(D)$ among the stellar binaries population, the conservative bound can then be obtained as
\begin{equation}
    \nu_\text{binary}=\int\limits_0^R\int\limits_0^\infty\int\limits_{r_\text{min}}^{r_C} \frac{d\nu}{dr_{\text{min}} dr_{\text{max}}} f(D) dr_\text{max}dDdr_\text{min}.
    \label{eq:nubinary}
\end{equation}

We infer the distribution $f(D)$ from measurements of binary orbital periods $P$ of nearby sun-like stars, with spectral types ranging from F6 to K3, by \cite{Raghavan}, who find a log-normal period distribution with $\log_{10} (P/\text{day})=5.03$ and $\sigma_{\log_{10}(P/\text{day})}=2.28$. 
Using Kepler's third law for circular orbits, the distribution of periods can be converted into the distribution of distances, $D$, which also follows a log-normal distribution $f(D)$. 
While it is known that the close-binary fraction is higher for metal-poor solar-type stars, such as those in UFDs \citep{Moe2019, Wyse_2020}, in the absence of a measured orbital period distribution we simply adopt the results for local (primarily solar metallicity) binary systems.

\begin{figure}[h!]
\includegraphics[width=1.\columnwidth]{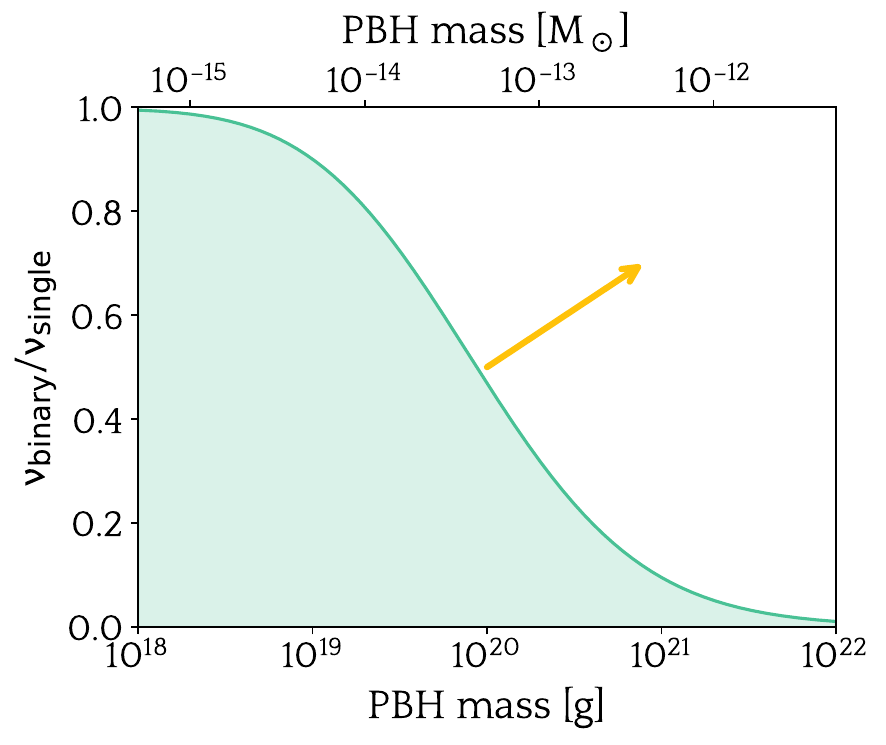}
\caption{\label{fig:single_to_binary} Ratio of the mean number of PBHs captured by stars in binary systems to the mean number captured by single stars, under the very conservative assumptions described in the text. While the coloured region represents a conservative estimate that cannot be reduced further, the arrow indicates that a more thorough analysis could potentially expand this region.}
\end{figure}

Assuming sun-like primary and companion stars, we computed the ratio $\nu_\text{binary}/\nu_\text{single}$ for various PBH masses\footnote{With Eq.~\eqref{eq:NPBHcap} and an LN distribution for $f(D)$, Eq.~\eqref{eq:nubinary} can be computed analytically in terms of Erf and Erfc functions.}. It is displayed on Fig. \ref{fig:single_to_binary}. We checked that this result changed only slightly with the stellar masses and the primary star's radius.

One can see that lighter PBHs are less affected by the presence of binaries, because the condition that they lose their energy sufficiently quickly (Eq.~\eqref{eq:rT}) implies that only the light PBHs with very small orbits, usually significantly smaller than the distance between the primary star and its companion, end up captured. For heavier PBHs, larger orbits are allowed, but these orbits are then perturbed and the PBHs may not end up being captured. However, we emphasise here that this result is a conservative lower bound on the number of captured PBHs in binaries. To determine the effects at large PBH masses, a full three-body study would be required.

We conclude from this calculation that most PBHs with masses $\lesssim 10^{19}$g are not affected by binary companions. Hence, we can safely use the expression \eqref{eq:nu} for $\nu(M)$ up to this mass. Above this value, the situation remains unclear and will need to be studied in the future. We note, however, that at a mass of $\sim 10^{20}$g, a sizeable fraction of $\sim 50 \%$ of PBHs is still unaffected by companion stars.

\end{appendix}

\end{document}